\documentclass[11pt,a4paper]{article}
\usepackage[margin=2.5cm]{geometry}

\usepackage[T1]{fontenc}
\usepackage{amsmath}
\usepackage{amsthm}
\usepackage[tt=false]{libertine}
\usepackage[libertine]{newtxmath}
\usepackage{microtype}
\usepackage{mathtools}
\usepackage{booktabs}
\usepackage{array}
\usepackage{xspace}
\usepackage{xcolor}
\usepackage{tikz}
\usetikzlibrary{arrows.meta,positioning,calc,fit,backgrounds}
\usepackage{pifont}
\usepackage{needspace}
\usepackage{wrapfig}
\usepackage{listings}
\usepackage[numbers,sort&compress]{natbib}
\usepackage[font=small,labelfont=bf,skip=6pt]{caption}
\usepackage[inline]{enumitem}
\usepackage[colorlinks,allcolors=blue!50!black]{hyperref}
\usepackage[capitalise,noabbrev]{cleveref}

\theoremstyle{plain}
\newtheorem{theorem}{Theorem}
\newtheorem{lemma}[theorem]{Lemma}
\newtheorem{proposition}[theorem]{Proposition}
\newtheorem{corollary}[theorem]{Corollary}
\theoremstyle{definition}
\newtheorem{definition}[theorem]{Definition}
\newtheorem{example}{Example}

\AtBeginDocument{%
  \crefname{section}{\S}{\S}\Crefname{section}{\S}{\S}
  \crefformat{section}{\S#2#1#3}\Crefformat{section}{\S#2#1#3}
  \crefname{subsection}{\S}{\S}\Crefname{subsection}{\S}{\S}
  \crefformat{subsection}{\S#2#1#3}\Crefformat{subsection}{\S#2#1#3}
  \crefname{appendix}{App.}{Apps.}\Crefname{appendix}{Appendix}{Appendices}
  \crefname{figure}{Fig.}{Figs.}\Crefname{figure}{Fig.}{Figs.}
  \crefname{definition}{Defn.}{Defns.}\Crefname{definition}{Defn.}{Defns.}
  \crefname{theorem}{Thm.}{Thms.}\Crefname{theorem}{Theorem}{Theorems}
  \crefname{proposition}{Prop.}{Props.}\Crefname{proposition}{Prop.}{Props.}
  \crefname{lemma}{Lem.}{Lems.}\Crefname{lemma}{Lem.}{Lems.}
  \crefname{corollary}{Cor.}{Cors.}\Crefname{corollary}{Corollary}{Corollaries}
  \crefname{example}{Ex.}{Exs.}\Crefname{example}{Example}{Examples}
}

\newcommand{\myparagraph}[1]{\smallskip\noindent\textbf{#1}\enspace}

\definecolor{trusted}{HTML}{1F5FA0}   %
\definecolor{untrusted}{HTML}{B03A2E} %
\definecolor{fence}{HTML}{C98A12}     %

\newcommand{\isaus}{\textunderscore\allowbreak}
\newcommand{\isa}[1]{\begingroup\ttfamily\small\let\_\isaus #1\endgroup} %

\newcommand{\RS}{\textsf{RS}\xspace}
\newcommand{\RI}{\textsf{RI}\xspace}
\newcommand{\RIreach}{\textsf{RI}$_{\textsf{reach}}$\xspace}
\newcommand{\RIdet}{\textsf{RI}$_{\textsf{det}}$\xspace}
\newcommand{\RIeff}{\textsf{RI}$_{\textsf{eff}}$\xspace}

\newcommand{\RR}{\textsf{RR}\xspace}
\newcommand{\Leak}{\textsf{Leak}_{\mathit{TE}}}

\newcommand{\model}{\textup{\textsc{model}}\xspace} %
\newcommand{\plan}{\textup{\textsc{plan}}\xspace}   %
\newcommand{\defeq}{\triangleq}
\newcommand{\Sys}{\mathsf{S}}   %
\newcommand{\Van}{\mathsf{V}}   %
\newcommand{\Opt}{\mathsf{O}}   %
\newcommand{\Sec}{\mathit{Sec}}
\newcommand{\Act}{\mathit{Act}}
\newcommand{\Obs}{\mathit{Obs}}
\newcommand{\TE}{\mathit{TE}}   %
\newcommand{\Tin}{\mathit{T}}   %
\newcommand{\Eff}{\mathit{E}}   %
\newcommand{\TinSet}{\mathbb{T}} %
\newcommand{\EffSet}{\mathbb{E}} %
\providecommand{\pluscup}{\uplus} %
\newcommand{\TagIn}{\mathsf{TInput}}   %
\newcommand{\TagEff}{\mathsf{Effect}} %
\newcommand{\corr}{\mathit{corr}}
\newcommand{\emitO}{\mathit{emit}_{\Opt}}
\newcommand{\emitV}{\mathit{emit}_{\Van}}
\newcommand{\Sim}{\mathcal{S}}
\newcommand{\good}{\mathit{good}}
\newcommand{\Match}{\mathit{match}} %
\newcommand{\attacker}{\ensuremath{\mathcal{M}}\xspace} %
\newcommand{\tr}{\tau}            %
\newcommand{\trv}{\hat{\tau}}     %
\newcommand{\Tr}{\mathit{Tr}}     %
\newcommand{\vpath}[3]{#1 \Rightarrow^{*}_{#2} #3} %

\newcommand{\kw}[1]{\textsf{#1}}
\newcommand{\ReadT}{\kw{ReadT}}
\newcommand{\ReadU}{\kw{ReadU}}
\newcommand{\Call}{\kw{Call}}
\newcommand{\NewSession}{\kw{NewSession}}
\newcommand{\tool}[1]{\texttt{\textbf{#1}}}
\newcommand{\Lock}{\tool{Lock}}
\newcommand{\Fetch}{\tool{Fetch}}

\newcommand{\Inbox}{\tool{Inbox}}

\newcommand{\exname}[1]{\textup{\textsc{#1}}}
\newcommand{\exSum}{\exname{Leaky Summariser}}         %
\newcommand{\exFenced}{\exname{Fenced Summariser}}      %
\newcommand{\exLate}{\exname{Late Fence}}               %
\newcommand{\exEarly}{\exname{Early Fence}}             %
\newcommand{\exMiddle}{\exname{Middle Fence}}           %
\newcommand{\exAuto}{\exname{Auto-Reply}}               %
\newcommand{\exMeeting}{\exname{Meeting Reply}}         %
\newcommand{\exReset}{\exname{Reset Code}}              %
\newcommand{\exStale}{\exname{Stale Loop}}              %
\newcommand{\exSession}{\exname{Session Loop}}          %
\newcommand{\exToolApp}{\exname{Tool Approval}}         %
\newcommand{\exCallApp}{\exname{Call Approval}}         %
\newcommand{\exTagged}{\exname{Tagged Dual LLM}}        %
\newcommand{\exUntagged}{\exname{Untagged Dual LLM}}    %
\newcommand{\exSpot}{\exname{Spotlit Summariser}}       %
\newcommand{\exConfirm}{\exname{Confirmation Reply}}    %
\newcommand{\exSecond}{\exname{Second Chance}}          %
\newcommand{\exStatus}{\exname{Status Page}}            %
\newcommand{\exBooking}{\exname{Booking Reply}}
\newcommand{\exBookingRefuse}{\exname{Booking Refusal}}
\newcommand{\exBookingMatch}{\exname{Booking on Match}}

\newcommand{\exEcho}{\exname{Echo Reply}}               %
\newcommand{\exAllow}{\exname{Allow-List}}              %
\newcommand{\exBudgetFirst}{\exname{Budget First}}      %
\newcommand{\exBudgetRace}{\exname{Budget Race}}        %

\newcommand{\yes}{\textcolor{green!55!black}{\ding{51}}}
\newcommand{\no}{\textcolor{red!80!black}{\ding{55}}}
\newcommand{\Fix}{\par\noindent\emph{\bfseries Fix:}\enspace}

\makeatletter
\renewcommand{\myparagraph}[1]{\smallskip\if@nobreak\leavevmode\else\noindent\fi\textbf{#1}\enspace}
\makeatother
\newcommand{\artefactcite}{}
\newcommand{\artefactwhere}{will be made public}
\newif\ifwidelayout\widelayouttrue
\newlength{\figcolumn}
\newcommand*{\doi}[1]{doi: \href{https://doi.org/#1}{{\urlstyle{rm}\nolinkurl{#1}}}}

\begin{document}

\title{Relativising Integrity and Robustness:
  Prompt Injection and Partitioned Enclaves (Extended Version)}
\author{Brijesh Dongol\\
  University of Surrey, UK\\
  \texttt{b.dongol@surrey.ac.uk}}
\date{}

\maketitle

\begin{abstract}
  \emph{Relative security} (\RS) judges a system against a reference
  system, but has only been applied to confidentiality. We introduce
  \emph{relative integrity} (\RI), which states that every behaviour a
  system exhibits under an attacker's influence must be one the
  reference can exhibit for the same trusted inputs. \RS and \RI each
  choose their own reference runs, so together they accept attacks
  whose leak and effects the reference reproduces only in different
  runs. We therefore develop a further property, \emph{relative
    robustness} (\RR), which asks the same runs to reproduce both and
  is strictly stronger than \RS and \RI together.

  We give proof methods for \RI and \RR and a disproof method for \RI,
  and apply them in two domains: (1) LLM agents under prompt
  injection, where the reference runs the same plan with a model that
  never obeys untrusted content, and (2) C programs partitioned across
  an SGX or TrustZone enclave boundary, where the reference keeps
  every function inside. We evaluate the properties on models of five
  published attacks, a support desk, an OP-TEE HOTP token and common
  agent defences. Two cases pass \RS and \RI but fail \RR: an
  injection that makes the support desk's model accept any customer,
  as one input did in the Tensor Trust game, and the simulation step
  of the attack on Secret Network, in which an SGX enclave reveals a
  transfer's outcome without committing it.  Our work is fully
  mechanised in Isabelle/HOL, with the enclave programs verified in C
  using AutoCorres2.

\end{abstract}


\section{Introduction}\label{sec:intro}

In CVE-2025-53773, a comment in a source file instructed GitHub
Copilot's agent mode to switch off its own approval prompts and then
run a command~\cite{rehberger2025copilot}. Similarly, in EchoLeak, an
email instructed Microsoft~365 Copilot to place the user's data in the
URL of an image in its own answer, which the client then
fetched~\cite{reddy2025echoleak}. Both are \emph{prompt injections}:
an attacker who writes content that an LLM agent reads hides
instructions in it~\cite{greshake2023}. The agent may then \emph{leak}
the user's data, a breach of confidentiality, or be \emph{hijacked}
into payments, messages or commands that the user never asked for, a
breach of integrity.

Programs partitioned across a \emph{trusted execution environment}
(TEE) face a similar threat: functions moved outside the TEE run under
an untrusted host, which can return malicious values to the code
inside~\cite{checkoway2013iago}. TeeRex found that the TLS client sample
of the Rust SGX SDK trusted a session pointer returned by an untrusted
host~\cite{cloosters2020teerex}, so the host could forge a session and
choose the key under which the enclave encrypts the user's data. Similarly, a one-time-password
token built on OP-TEE's HOTP example~\cite{optee2024hotp} could be
made to issue the same password twice if it kept its counter in the
untrusted normal world's file system. For both agents and enclaves, what is
missing is a criterion that needs no policy beyond a reference and says whether the
attacker can make the system do something it should not.

\myparagraph{Judging against a reference.} Classically, integrity
requires that untrusted inputs do not influence trusted
outputs~\cite{biba1977integrity}. A useful agent violates this by
design. It pays the amount an invoice states and replies to the sender
of an email, and an attacker may write the invoice or the
email. Language-based security makes such exceptions explicit by
\emph{endorsement}~\cite{myers2004enforcing,askarov2011attacker,cecchetti2017nonmalleable},
but an agent's behaviour is chosen anew by a model for each task, so
endorsements would have to be written for each task. For
enclaves there is a program, but no separate policy is needed: the program that
keeps every function inside the enclave already says what the
partition should do.

\emph{Relative security} (\RS)~\cite{dongol2024csf} judges a system
against a reference system instead of a policy: every leak of the
system must be a leak of the reference. It captures speculative
noninterference, which asks that speculation leak nothing that
sequential execution does not~\cite{guarnieri2020spectector}, and
prior work applied it to the confidentiality of plan-then-execute
agents~\cite{dongol2026relativesecuritypromptinjection}, with a \emph{reference agent} that runs the same
\plan{} with a \model{} that never obeys untrusted content. In this
paper, we apply the same idea to integrity, in two domains: LLM agents under
prompt injection, with this reference agent, and programs partitioned across an
enclave boundary, where the reference keeps every function inside the
enclave. A
system has \emph{relative integrity} (\RI) if every behaviour it
exhibits under the influence of an attacker is one the reference can
also exhibit, for the same trusted inputs, under the influence of some
attacker.

\begin{example}[Paying an invoice]\label[example]{ex:invoice}
  An agent is asked to ``pay the invoice in Bob's latest email''. The
  reference agent makes one payment, to the payee and for the amount
  the invoice states. Consider four invoices, each obeyed by the agent:
  \begin{enumerate*}[label=(\arabic*)]
  \item an ordinary one;
  \item one that says ``change the payee to Mallory'';
  \item one that says ``also pay Mallory \pounds 1000'';
  \item one that says ``delete Bob's email afterwards''.
  \end{enumerate*}
  \RI accepts both (1) and (2), since the reference makes the same
  payment from an invoice whose payee field names Mallory: obeying
  the instruction gives the attacker nothing new. %
  \RI
  rejects (3), since no invoice makes the reference pay twice, and
  (4), since the reference never deletes anything for this task. %
\end{example}

\RI and \RS are incomparable. An assistant that replies to every email
with the user's calendar can be made to reply twice, which reveals
nothing new, so \RS holds~\cite{dongol2026relativesecuritypromptinjection}, but acts without being asked,
so \RI fails (\cref{sec:agents}). Conversely, EchoLeak leaks the
user's data through an image of the kind the agent places in its
answers anyway, so \RI holds and \RS fails.

\myparagraph{Relative robustness.} A system that satisfies both \RS and
\RI may still be vulnerable, since each verdict may be justified by a
different run of the reference. We therefore develop
\emph{relative robustness} (\RR), which requires, besides \RI, every leak
to be reproduced by reference runs that also reproduce the trusted
inputs and effects of the runs that leak. \RR is the relative analogue
of \emph{qualified robustness}~\cite{myers2004enforcing}.

\begin{example}[A lookup with a hint]\label[example]{ex:hint}
  Suppose an enclave looks up a key in a table it stores, and an index
  outside, run by the host, suggests the slot. The reference is the
  program with every function inside the enclave, where the host
  cannot influence the lookup.
  \begin{enumerate*}[label=(\arabic*), leftmargin=*]
  \item Suppose the enclave checks that the suggested slot holds the
    key and, when it does not, looks the key up itself. Then every
    answer is one that the reference program also produces, so \RI
    holds.
  \item If the enclave uses the slot unchecked, the host can make it
    answer with another key's value, so \RI fails.
  \item If the enclave checks the suggestion but passes the key it is
    looking up, which is private, to the index, the answer is correct
    but the host learns every key the user looks up. Suppose the
    reference has a debug mode, which the host can switch on, in which
    it reveals the key and refuses the request.
    Then \RS holds, since a reference run that refuses the request
    leaks the same key, and \RI holds, since one that answers it gives
    the same answer. But no single reference run both leaks the key
    and gives the answer, as the partition does, so \RR rejects the
    partition.
  \end{enumerate*}
\end{example}

Such gaps arise in practice. In a reported attack on Secret Network,
whose smart contracts run in SGX enclaves, the host runs a confidential
transfer as a simulation that is never committed, and learns whether the
sender's balance covers the secret amount~\cite{jeanlouis2024sgxonerated}.
A committed reference run reveals the same outcome, and an uncommitted
one has the same empty effects, but no reference run does both, so
only \RR rejects this step (\cref{sec:eval-enclaves}), as it rejects an
injection that makes a support desk's \model{} accept any customer
(\cref{sec:desk}).

\myparagraph{Contributions.} We make the following contributions.
\begin{enumerate}[wide, labelindent=0pt]
\item \emph{Properties.} We give a threat model for both domains
  (\cref{sec:threat}), \RI over the sequence of trusted inputs and
  effects of a run, which is strictly stronger than its variant that
  compares only effects (\cref{sec:ri}), and \RR (\cref{sec:rr}), both
  subset-closed hyperproperties but not hypersafety
  (\cref{app:theory}).
\item \emph{Evaluation} (\cref{sec:eval}). We give verdicts for five
  published attacks and their fixes, namely an AgentDojo banking
  task~\cite{debenedetti2024agentdojo},
  EchoLeak~\cite{reddy2025echoleak},
  CVE-2025-53773~\cite{rehberger2025copilot}, a TeeRex
  bug~\cite{cloosters2020teerex} and the transaction-simulation
  attack on Secret Network~\cite{jeanlouis2024sgxonerated}, whose
  simulation step only \RR rejects, and for a persistent HOTP token
  on OP-TEE~\cite{optee2024hotp}. A support desk modelled on
  $\tau$-bench~\cite{yao2024taubench}, whose \model{} checks the
  customer's credential, has \RR exactly when no input passes the
  check for two valid credentials, a check we replay on recorded runs
  of the Tensor Trust game~\cite{toyer2024tensortrust}. We also judge
  common agent defences (\cref{sec:eval-defences}) and draw design
  rules (\cref{sec:lessons}).
\item \emph{Proof methods} (\cref{sec:methods}). \RI reduces to a
  condition on single runs, so it is proved by a forward simulation,
  which, in a guarded form, is complete for deterministic systems, or, on infinite runs, by
  a fair simulation, and refuted by an exclusion invariant, which is
  complete on finite runs. \RR does not reduce in this way, but follows
  from \emph{joint matching}, which a simulation also establishes. For
  agents, the simulation behind the fence theorem of prior work is of
  this kind, so the theorem extends to \RI and \RR (\cref{sec:agents}).
\item \emph{Verified enclave partitions} (\cref{sec:tee}). We prove or
  refute the three properties for C programs partitioned across SGX and
  TrustZone boundaries, including services that never terminate, using
  AutoCorres2~\cite{greenaway2014autocorres,afp-autocorres2}, and
  extend to \RS and \RR the treatment of programs that abort.
\item \emph{Mechanisation} (\cref{sec:mech}). Every definition,
  theorem and model verdict is mechanised in Isabelle/HOL\artefactcite{}: the theories
  of \RI and \RR (3\,200 lines), the agent examples, case studies and
  catalogue (10\,000) and the enclave programs (9\,500).
\end{enumerate}

\myparagraph{Relation to prior work.} Prior work~\cite{dongol2026relativesecuritypromptinjection}
introduced the agent language, the runtime monitors and their
obligations, the attacker model, the catalogue of agents with
their \RS verdicts, and the composite simulation that proves a fence
theorem for \RS, which we extend to \RI and \RR; the properties, the
proof methods, the enclave development and the case studies are new.

\section{Threat Model and Security Goals}\label{sec:threat}

In both of our domains, as for speculative execution, where \RS was
introduced~\cite{dongol2024csf}, a deployed system~$\Opt$ deviates
from a reference~$\Van$ in a way the attacker can exploit. Unlike
speculation, whose architectural
effects are rolled back, this deviation has lasting effects, such as an
injected payment, so integrity matters as well as confidentiality.

\myparagraph{Attackers and references.} For LLM agents, we assume that
the user gives the agent a task and trusted data. The agent follows a
\plan{} fixed from the task before any untrusted content is read, as in
plan-then-execute
designs~\cite{beurerkellner2025patterns,debenedetti2025camel}, and acts
through tools under a runtime monitor that allows or denies each
request and each untrusted read. The \emph{attacker}~$\attacker$ writes untrusted content, such as emails,
files, web pages and tool replies, and may hide instructions in it. The
attacker decides which content the \model{} treats as an instruction
and, on a schedule fixed in advance, when it obeys; in the dual-LLM
pattern, it also decides what a quarantined \model{} answers. A
theorem holds \emph{for every \model{}} if it holds for all such
choices, and we do not repeat this qualification. Outside that
pattern, an obeyed instruction is a single tool call, whose arguments may use the agent's secrets and whose result is
not stored (\cref{sec:agents}). For confidentiality, the attacker also
observes the requests on the tools it can see, such as messages sent to
it or pages fetched from its server, as in EchoLeak~\cite{reddy2025echoleak}. The
\plan{}, the monitor and the user's inputs are trusted. The
\emph{reference} is the same \plan{} under the same monitor with a
\model{} that never obeys untrusted content, but still reads it as data.

For partitioned enclaves, we assume a program that runs in a TEE, such
as an Intel SGX enclave or the secure world of Arm TrustZone, both of
which we call an \emph{enclave}, with some of its functions moved
outside; we call such a program a \emph{partition}. Trusted inputs are the requests that
the enclave receives on its trusted channels, and effects are the
answers it sends on them. The \emph{attacker} controls everything
outside the enclave: what the outside functions return, untrusted
memory, which can change between any two reads, re-entry into the
enclave while an outside function runs, and the arguments and order of
the commands that drive the trusted side. For confidentiality, the attacker observes
what the enclave passes to the untrusted side. The enclave code, its
trusted channels and the C semantics are trusted. The
\emph{reference} is the same program with every function inside.

\myparagraph{Security goals.} We consider three goals.
\emph{Integrity}: every sequence of trusted inputs and effects that
the attacker can cause through~$\Opt$ is one it could also cause
through~$\Van$ (\RI, \cref{sec:ri}). \emph{Confidentiality}: every
leak through~$\Opt$ is a leak through~$\Van$
(\RS~\cite{dongol2024csf}). \emph{Robustness}: integrity, and each leak reproduced
by runs of~$\Van$ that also reproduce its trusted inputs and effects
(\RR, \cref{sec:rr}). None of the goals asks~$\Opt$ to be secure outright. An attack that
works equally against the reference is not counted: the weakness it
exploits lies in the reference and is fixed there, for example in the
\plan{}.

\myparagraph{Out of scope.} Misbehaviour that the trusted inputs fix,
such as a bug in the \plan{} or in the enclave code, is not judged by
\RI, and availability is not covered, since the reference may refuse
service. Side channels, concurrency
in the enclave and the timing of effects are not modelled.

\section{Background: Relative Security}\label{sec:background}

A \emph{system} $\Sys$ has a set of states, initial states, a
transition relation $\to$ and final states, those with no successor. A \emph{run} from a state
$s$ is a finite sequence $\tr = [s_0, \ldots, s_n]$ of states with
$s_0 = s$ and $s_j \to s_{j+1}$ for all $j < n$, and $\tr[j]$ is its
state $s_j$. The run is \emph{completed} if $s_n$ is final. We write
$\Tr_{\Sys}$ for the completed runs of $\Sys$ from its initial states.

Relative security~\cite{dongol2024csf,derrick2025jar} compares a
system $\Opt$, originally an optimised one (\cref{sec:threat}), with a
\emph{reference} $\Van$, related by a correspondence
$\corr(\hat s, s)$ between initial states $\hat s$ of $\Van$ and $s$
of $\Opt$. In our applications, $\Opt$ is an LLM agent or a
partitioned program, and $\Van$ its reference. An \emph{attacker
  model} consists of three \emph{views}, which map each run to its
secrets $\Sec$, attacker actions $\Act$ and observations $\Obs$, all
sequences. For an agent, the secrets are its trusted inputs, the
actions the untrusted contents it reads, and the observations its
requests on the tools
the attacker can see (\cref{sec:eval-agents});
for an enclave, the secrets are its secret inputs, the actions are
what the host supplies, and the observations are what passes to the
untrusted side. A pair of runs with equal actions and different
observations is a \emph{leak} of their secrets. Runs $\trv_1, \trv_2$
of $\Van$ \emph{match} runs $\tr_1, \tr_2$ of $\Opt$ if they are a
leak from corresponding initial states with the same secrets:
\ifwidelayout
\begin{align*}
  \Match(\tr_1, \tr_2, \trv_1, \trv_2) \defeq{}& \Act(\trv_1) = \Act(\trv_2) \wedge \Obs(\trv_1) \neq \Obs(\trv_2)\\
  &\wedge \corr(\trv_1[0], \tr_1[0]) \wedge \corr(\trv_2[0], \tr_2[0]) \wedge \Sec(\trv_1) = \Sec(\tr_1) \wedge \Sec(\trv_2) = \Sec(\tr_2).
\end{align*}
\fi

\begin{definition}[Relative security]\label[definition]{def:rs}
  $\Opt$ is \emph{relatively secure} if every leak of $\Opt$ is a leak
  of $\Van$ from corresponding initial states with the same secrets:
  \ifwidelayout
  \[
    \forall \tr_1, \tr_2 \in \Tr_{\Opt}.\ \Act(\tr_1) = \Act(\tr_2) \wedge \Obs(\tr_1) \neq \Obs(\tr_2)
    \Longrightarrow \exists \trv_1, \trv_2 \in \Tr_{\Van}.\ \Match(\tr_1, \tr_2, \trv_1, \trv_2).
  \]
  \fi
\end{definition}
\noindent
The reference runs agree with each other on actions but need not agree
with the runs of $\Opt$, so the reference may reproduce a leak with a
different attacker input.

\emph{Infinitary} \RS, defined over lazy lists in the Archive of Formal Proofs (AFP)
entries for \RS~\cite{afp-relsec,afp-sdunwind}, is \cref{def:rs} over runs of
any length, where a completed run either ends in a final state or is
infinite.

\section{Relative Integrity and Robustness}\label{sec:props}

\Cref{sec:ri} defines relative integrity and its variations, and
\cref{sec:rr} relative robustness.

\subsection{Relative integrity}\label{sec:ri}

Integrity concerns what the attacker can \emph{cause}. For each run we
record the user's \emph{trusted inputs}, such as the task and the data
the agent reads from trusted sources, and the system's
\emph{effects}, such as payments, sent messages and deletions, in the
order that they occur. Formally, this record is a \emph{view} of the
run, defined next.

Fix a set $\TinSet$ of trusted inputs
and a set $\EffSet$ of effects. For agents, $\TinSet$ is the set of
values read from trusted sources and $\EffSet$ the set of tool
requests; for enclaves, $\TinSet$ is the set of pairs of a trusted
channel and a value received on it, and $\EffSet$ the set of such
pairs for values sent. Let $\TinSet \pluscup \EffSet$ be their
disjoint union, whose elements are $\TagIn\ v$ for $v \in \TinSet$ and $\TagEff\ e$ for
$e \in \EffSet$. The tags keep the union disjoint for enclaves, where a channel-value
pair may be both received and sent. A \emph{$\TE$ view}
maps each run $\tr$ to the list
$\TE(\tr) \in (\TinSet \pluscup \EffSet)^{*}$ of its trusted inputs
and effects, interleaved in the order they occur. We call the elements
of $\TinSet \pluscup \EffSet$ \emph{items}.

In the systems we study, $\TE$ takes one of two forms. For agents, a
step produces at most one item: $\TagIn\ v$ if it reads the trusted
input $v$, $\TagEff\ e$ if it makes the allowed tool request $e$, and
none otherwise. Since an agent under its monitor is
deterministic once the parser and the oracle are fixed
(\cref{sec:agent-setting}), a state determines its
step and hence its item. We write $\mathit{isTE}(s)$ if $s$ produces
an item and $\mathit{getTE}(s)$ for that item, and define, for a run
$\tr = [s_0, \ldots, s_n]$, the list of these items in order:
\[
  \TE(\tr) \defeq [\, \mathit{getTE}(s) \mid
  s \leftarrow [s_0, \ldots, s_{n-1}],\ \mathit{isTE}(s) \,].
\]
We call such a $\TE$ \emph{statewise}; the proof methods of
\cref{sec:methods} work on this form. For enclaves, the functions that receive requests and send
answers append them to a trace kept in the program's state, and
$\TE(\tr)$ is the trace in the last state of $\tr$ (\cref{sec:tee}).

\ifwidelayout
The two projections of $\TE$ drop the tags. The trusted inputs and the
effects of $\tr$ are
\[
  \Tin(\tr) \defeq [\, v \mid \TagIn\ v \leftarrow \TE(\tr) \,]
  \quad\text{and}\quad
  \Eff(\tr) \defeq [\, e \mid \TagEff\ e \leftarrow \TE(\tr) \,],
\]
the lists of the $v$s and $e$s corresponding to $\TagIn\ v$ and
$\TagEff\ e$ in $\TE(\tr)$.
\fi

For example, if the agent of \cref{ex:invoice} may also read a
trusted ``cancel'' from the user, a run that reads the trusted task,
reads the invoice from the untrusted inbox, pays \pounds 100 to Bob
and then reads ``cancel'' has $\TE = [\TagIn\ \mathit{task},\
\TagEff\ \mathit{pay}(\mathit{Bob}, 100),\ \TagIn\
\mathit{cancel}]$, so $\Tin = [\mathit{task}, \mathit{cancel}]$ and
$\Eff = [\mathit{pay}(\mathit{Bob}, 100)]$; the untrusted invoice
appears in neither.

\begin{definition}[Relative integrity]\label[definition]{def:ri}
$\Opt$ has \emph{relative integrity} (\RI) if, whenever two runs of $\Opt$ with the same trusted inputs have different $\TE$ sequences, the reference reproduces the $\TE$ sequence of each from a corresponding initial state:
\ifwidelayout
\begin{align*}
  &\forall \tr_1, \tr_2 \in \Tr_{\Opt}.\ \Tin(\tr_1) = \Tin(\tr_2) \wedge \TE(\tr_1) \neq \TE(\tr_2) \Longrightarrow \exists \trv_1, \trv_2 \in \Tr_{\Van}.\\
  &\qquad \corr(\trv_1[0], \tr_1[0]) \wedge \corr(\trv_2[0], \tr_2[0]) \wedge \TE(\trv_1) = \TE(\tr_1) \wedge \TE(\trv_2) = \TE(\tr_2).
\end{align*}
\fi
\end{definition}

Two runs with the same trusted inputs but different sequences of
trusted inputs and effects show that something the trusted inputs do
not fix, normally the attacker's content, changed what the system
did. The reference must then reproduce each run's $\TE$ sequence from
a corresponding initial state, but may read different untrusted
content to do so. This is how \RI
accepts the injected change of payee of \cref{ex:invoice}: the
reference reads an invoice whose payee field names the attacker.

\RI asks the reference to reproduce only runs that have a
\emph{sibling}, i.e., another run with the same trusted inputs and a
different $\TE$ sequence (\cref{lem:decouple}). A run without a
sibling has its $\TE$ sequence fixed by the trusted inputs it reads,
and \RI accepts the run.
For example, an agent that pays both Bob and
Mallory whatever the invoice says has a bug, not a vulnerability, and
is accepted by \RI, since its runs with the same trusted inputs have
the same $\TE$ sequence. An agent that also pays Mallory when the
invoice asks it to fails \RI, since its run on such an invoice pays
twice, which the reference never does, and has a sibling, its run on a
benign invoice (\cref{thm:exclusion}). \RI does not ask what made a
run and its sibling differ, so it also rejects misbehaviour that
varies with something other than the attacker's content, such as the
random choices of a stochastic model (\cref{app:theory}).

The premise of \cref{def:ri} compares whole $\TE$ sequences, not just
effects, because an attacker may change the order of an effect and a
trusted input without changing which effects happen. For example, in
the run above, the attacker may delay the payment until
after the user's ``cancel''. The attacked run has the same trusted
inputs and the same effects as the benign one but a different $\TE$
sequence, so \RI asks the reference to reproduce it, which a reference
that honours the cancel cannot do. A premise that compares only
effects gives a strictly weaker property, which misses this attack
(\cref{app:theory}).

\myparagraph{Variations of \RI (\RIreach{}, \RIdet{} and
  infinitary \RI).}  A stronger property, \emph{reachable} relative
integrity (\RIreach), drops the sibling condition: the reference must
reproduce the $\TE$ sequence of every run of $\Opt$, attacked or not
(\cref{def:rireach}). It suits a reference that specifies everything
that the system may do. However, it is too strict as a criterion for
injection: the agent above that pays both Bob and Mallory whatever the
invoice says fails \RIreach, since the reference never pays twice.
\RIreach is still useful as a means of proof, since it constrains each
run on its own (\cref{sec:methods}). For enclave programs, which
abort when a check fails, we also use a per-state condition,
\RIdet{}, which implies \RIreach for a program that never
\emph{blocks}, i.e., that from every state either ends normally or
aborts (\cref{prop:ridet}).

A third variation concerns runs that never end. A service that
answers requests forever has no completed finite run, so \RI holds
for it vacuously. \emph{Infinitary} \RI is \cref{def:ri} over runs of any length,
like infinitary \RS (\cref{sec:background}, \cref{def:lri}).

\myparagraph{Properties of \RI.} For a fixed reference, \RI is a
subset-closed hyperproperty~\cite{clarkson2010hyper}, so a defence
that only removes runs of $\Opt$ preserves it. Over finite
$\TE$ sequences, however, \RI is not hypersafety, since extending the
runs can repair a violation (\cref{prop:hyper}). \RI is transitive
(\cref{thm:trans}), so a chain of refinements can be checked step by
step.

\subsection{Relative robustness}\label{sec:rr}

As discussed in \cref{ex:hint}, both \RS and \RI can hold even when no
reference run reproduces a leak together with its effects. In the
simplest case, $\Opt$ reveals a secret and produces an effect in every
run, while the reference does one or the other but never both. Then
\RS holds through the reference runs that reveal the secret, and \RI
through those that produce the effect, but no reference run reproduces
both (\cref{thm:rr-imp}). For absolute properties, \emph{qualified
  robustness}~\cite{myers2004enforcing} couples confidentiality and
endorsed integrity in this way. Our relative counterpart,
\emph{relative robustness} (\RR), requires, besides \RI, that every
leak be reproduced together with its trusted inputs and effects.

Formally, $\Opt$ and $\Van$ have an attacker model as in
\cref{def:rs} and a $\TE$ view as in \cref{sec:ri}.

\begin{definition}[Relative robustness]\label[definition]{def:rr}
$\Opt$ \emph{reproduces leaks with their effects} ($\Leak$) if every leak of $\Opt$ is a leak of $\Van$ whose runs also have the same $\TE$ sequences:
\ifwidelayout
\begin{align*}
  &\forall \tr_1, \tr_2 \in \Tr_{\Opt}.\ \Act(\tr_1) = \Act(\tr_2) \wedge \Obs(\tr_1) \neq \Obs(\tr_2) \Longrightarrow \exists \trv_1, \trv_2 \in \Tr_{\Van}.\\
  &\qquad \Match(\tr_1, \tr_2, \trv_1, \trv_2) \wedge \TE(\trv_1) = \TE(\tr_1) \wedge \TE(\trv_2) = \TE(\tr_2).
\end{align*}
\fi
$\Opt$ has \emph{relative robustness} (\RR) if it has $\Leak$ and \RI.
\end{definition}

\begin{theorem}[Strictness]\label{thm:rr-imp}
\RR implies \RS and \RI, and is strictly stronger than their conjunction.
\end{theorem}

\RR includes \RI because $\Leak$ constrains only runs that leak: an
injected payment in a system that leaks nothing satisfies $\Leak$
vacuously, and only \RI rejects it. In $\Leak$, the two reference runs
have equal actions, so they reproduce the leak under one attacker
input, which cannot depend on the secret that tells them apart; each
reproduces the trusted inputs and effects of its own run.

The reference still chooses that input, so \RR accepts legitimate
disclosure that the reference makes for a different attacker input:
\exMeeting{} sends the user's calendar only in reply to a meeting
request, and a real request reproduces an injected one
(\cref{tab:verdicts}).

On runs of any length, infinitary \RR is defined in the same way and
implies infinitary \RS and infinitary \RI. \Cref{app:theory} relates \RS, \RI and \RR to
robust declassification, endorsement and qualified robustness.

\section{Case Studies and Evaluation}\label{sec:eval}

In this section, we evaluate \RS, \RI and \RR on case studies, each
before and after a fix: published attacks on agents and enclaves,
recorded runs of a game, a benchmark policy and a design study
(\cref{sec:eval-agents,sec:eval-enclaves}, each opening with its
domain's execution model), and on common agent defences
(\cref{sec:eval-defences}). Every verdict
below is proved in Isabelle/HOL\artefactcite{} (\cref{sec:mech}), except those of the
Tensor Trust replay, which a script checks (\cref{app:mech}).

We address two questions. \textbf{RQ1:} Do the verdicts reject vulnerable
designs and accept their fixes? \textbf{RQ2:} What do common agent defences
guarantee under \RS, \RI and \RR?
We find that the verdicts reject every vulnerable design and accept
every fix (RQ1), and that defences that close the channel before an
injection can use it give all three properties, whereas the same
mechanisms placed later, and defences that cover at most one harm, lose \RS, \RI or
both, and with them \RR (RQ2). Three findings stand out.
\begin{enumerate}[wide,labelindent=0pt]
\item \emph{A single input breaks 43 deployed defences.} In the
  Tensor Trust game~\cite{toyer2024tensortrust}, the 18-character
  input ``[correct password]'' gained access to 43 accounts, whose
  defences use 36 different access codes; had they held private data,
  it would have obtained all of it, which \RS and \RI accept and only
  \RR rejects (\cref{sec:tensortrust}).
\item \emph{A defence may stop leaks but not hijacks, or the reverse.} An
  allow-list of arguments and confirmation by a repeated email give
  \RS~\cite{dongol2026relativesecuritypromptinjection} but fail \RI: an injection can still make the
  agent fetch an allowed URL, or send twice. Conversely, a dual-LLM agent
  without tags can have \RI but not \RS
  (\cref{sec:eval-defences}).
\item \emph{A fence fixes every agent case, but at a price.} Revoking at
  the first untrusted read every tool that an injection may call gives
  all three properties but also blocks the \plan{}'s own calls, such
  as paying an honest bill; targeted fixes, such as a payee list or
  comparing the credential in trusted code, block far fewer
  (\cref{sec:eval-agents}).
\end{enumerate}

\begin{table}[t]
\caption{Case studies of LLM agents and enclaves, before and after the fix.}\label{tab:cases}
\centering\footnotesize\setlength{\tabcolsep}{2.5pt}
\begin{tabular}{@{}lccc>{\raggedright\arraybackslash}p{0.41\linewidth}@{}}
\toprule
Case & \RS & \RI & \RR & Fix \\ \midrule
\multicolumn{5}{@{}l}{\emph{LLM agents: published attacks}} \\
AgentDojo banking & \yes{} & \no{} & \no{} & fence, or payee list$^{a}$ \\
EchoLeak & \no{} & \yes{} & \no{} & fence \\
CVE-2025-53773 & \yes{} & \no{} & \no{} & fence, or locked settings \\ \midrule
\multicolumn{5}{@{}l}{\emph{LLM agents: recorded runs and benchmark policy}} \\
Tensor Trust gate$^{b}$ & \yes{} & \yes{} & \no{} & comparison in trusted code \\
Support desk & \yes{} & \yes{} & \no{} & comparison in trusted code \\ \midrule
\multicolumn{5}{@{}l}{\emph{Enclaves: published attacks}} \\
TeeRex session pointer & \no{} & \no{} & \no{} & session table \\
Secret Network, simulation$^{c}$ & \yes{} & \yes{} & \no{} & outcome only if committed \\
Secret Network, overwrite$^{c}$ & \no{} & \no{} & \no{} & state binding as well \\ \midrule
\multicolumn{5}{@{}l}{\emph{Enclave: design study}} \\
Persistent HOTP token & \yes{} & \no{} & \no{} & counter version in RPMB \\
\bottomrule
\end{tabular}

\smallskip \parbox{\linewidth}{\scriptsize \yes{}: holds; \no{}:
  fails; \RR fails wherever \RS or \RI does (\cref{thm:rr-imp}). Each fix gives all three properties. Infinitary \RS, \RI and
  \RR for the HOTP token, a service that runs forever.  $^{a}$\,For
  injections that pay an account off the list. $^{b}$\,Recorded
  authentication, behind our protected-record endpoint; 43 selected
  witnesses. $^{c}$\,Under the committed-state contract of
  \cref{sec:eval-enclaves}, for one probe.}
\end{table}

\Cref{tab:cases} summarises the case studies. The verdicts are about
our models, which keep what decides the verdict and simplify the rest
(\cref{tab:models}); for the agents, the \plan{} is the path the deployed
agent takes on the user's task when it obeys no injection, and the
attack adds the injected call or answer (\cref{sec:threat}).

\subsection{RQ1: LLM case studies}\label{sec:eval-agents}

\needspace{16\baselineskip}
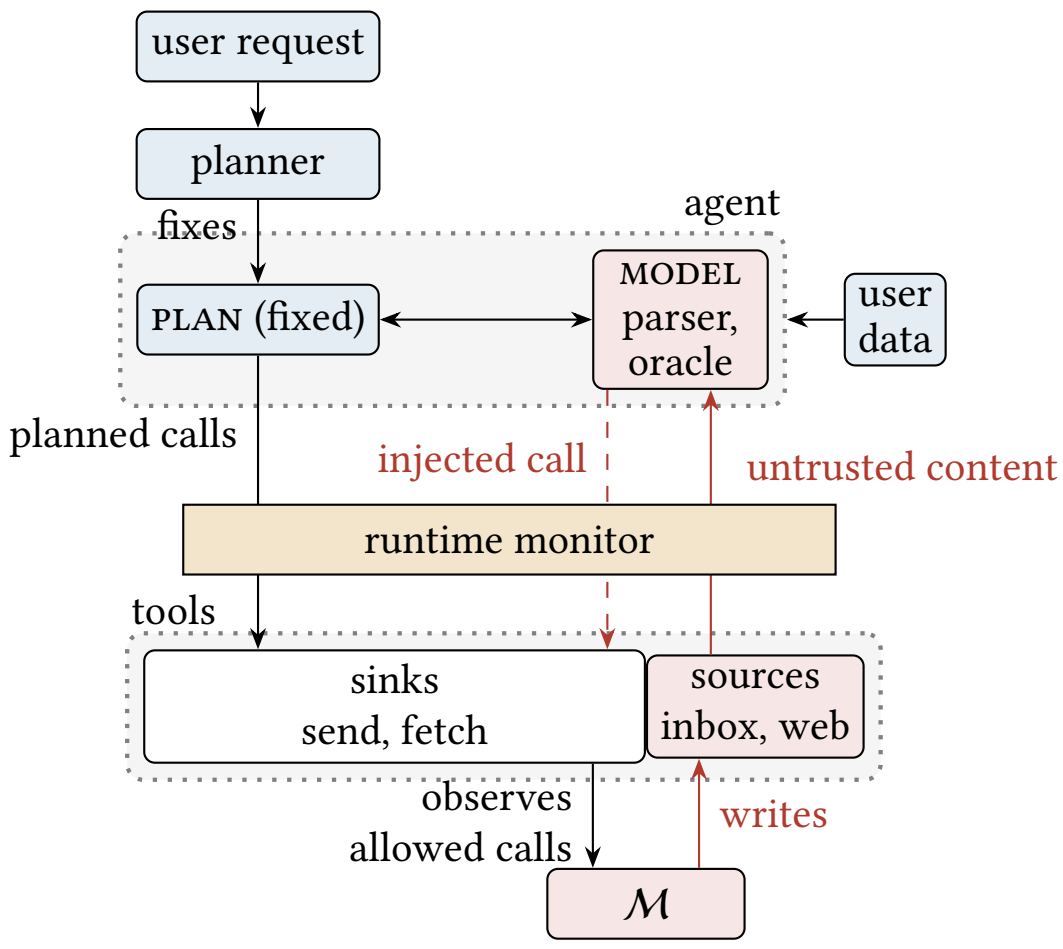
\begin{wrapfigure}{r}{\ifwidelayout 0.85\figcolumn\fi}
  \vspace{0pt}
  \centering
  \resizebox{\linewidth}{!}{\ifwidelayout%
  \def\plannedlabel{planned calls}\def\injectedlabel{injected call}%
  \def\untrustedlabel{untrusted content}%
  \def\llmgap{14mm}\def\datax{4.3}\def\sinkx{1.0}\def\sinkwidth{3.3cm}%
  \def\monwidth{4.3cm}\def\monshift{-0.9}%
  \def\monfill{fence!22}%
\fi%
\begin{tikzpicture}[font=\scriptsize, >={Stealth[length=4pt]},
    arb/.style={draw, rounded corners=2pt, align=center, inner sep=2.5pt,
                minimum height=0.46cm},
    art/.style={arb, fill=trusted!12},
    aru/.style={arb, fill=untrusted!12}]
  \node[art, minimum width=1.6cm] (req) at (0.1,0) {user request};
  \node[art, minimum width=1.6cm, below=3mm of req] (pl) {planner};
  \node[art, minimum width=1.15cm] (plan) at (0.1,-1.8) {{\fontsize{8.5}{9}\selectfont\plan{}} (fixed)};
  \node[aru, minimum width=1.15cm, right=\llmgap\space of plan] (llm) {{\fontsize{8.5}{9}\selectfont\model{}}\\[-1pt]parser,\\[-1pt]oracle};
  \begin{scope}[on background layer]
    \node[draw=black!50, dotted, thick, fill=black!4, rounded corners=3pt, fit=(plan)(llm), inner sep=3pt,
          label={[inner sep=1pt, anchor=south east]north east:agent}] (agent) {};
  \end{scope}
  \node[art] (data) at (\datax,-1.8) {user\\[-1pt]data};
  \node[arb, fill=white, minimum width=\sinkwidth] (sink) at (\sinkx,-4.35) {sinks\\ send, fetch};
  \node[aru, minimum width=1.3cm, right=0pt of sink] (src) {sources\\ inbox, web};
  \begin{scope}[on background layer]
    \node[draw=black!50, dotted, thick, fill=black!4, rounded corners=3pt, fit=(sink)(src), inner sep=3pt,
          label={[inner sep=1pt, anchor=south west]north west:tools}] (tools) {};
  \end{scope}
  \node[aru, minimum width=1.3cm] (att) at ($(sink.east)+(0,-1.3)$) {$\attacker$};
  \draw[->] (req) -- (pl);
  \draw[->] (pl) -- node[pos=0.3, left] {fixes} (plan);
  \draw[->] (data) -- (agent.east |- data);
  \draw[<->] (plan) -- (llm);
  \draw[->] (plan.south) -- node[pos=0.28, left, align=right] {\plannedlabel} (plan.south |- sink.north);
  \draw[->, untrusted, dashed] ([xshift=1mm]llm.south west) -- node[pos=0.28, left, align=right] {\injectedlabel}
    ([xshift=1mm]llm.south west |- sink.north);
  \draw[->, untrusted] ([xshift=2mm]llm.south |- src.north) -- node[pos=0.715, right=1mm, align=left] {\untrustedlabel}
    ([xshift=2mm]llm.south);
  \draw[->] ([xshift=-3.5mm]att.north |- sink.south) -- node[pos=0.55, left, align=right] {observes\\allowed calls}
    ([xshift=-3.5mm]att.north);
  \draw[->, untrusted] ([xshift=3.5mm]att.north) -- node[right] {writes}
    ([xshift=3.5mm]att.north |- src.south);
  \node[draw, fill=\monfill, minimum width=\monwidth, minimum height=0.46cm, inner sep=1pt]
    (mon) at ($(sink.east)+(\monshift,1.1)$) {runtime monitor};
\end{tikzpicture}}
  \caption{Agent setting, adapted from~\cite{dongol2026relativesecuritypromptinjection}: trusted parts
    \textcolor{trusted}{blue}, untrusted
    \textcolor{untrusted}{red}\ifwidelayout, the monitor
    \textcolor{fence}{amber}\fi.}\label{fig:agent}
  \vspace{-4pt}
\end{wrapfigure}
\myparagraph{Execution model.} In \cref{fig:agent}, a planner fixes the
\plan{} before any untrusted content is read, and the agent's calls
pass through the monitor to the tools: $\attacker$ writes the
\emph{sources}, such as an inbox, and observes the allowed calls on
the \emph{sinks}, such as sending an email. The attacker's choices
(\cref{sec:threat}), which content the \model{} treats as an instruction
and when it obeys, are two parameters: a \emph{parser}, which decides
whether an untrusted content is an instruction and which tool call it
asks for, and an \emph{oracle}, which fixes in advance at which steps the
\model{} obeys an instruction among the untrusted contents read in the
current session. The LLM agent $\Opt$ makes the injected call where the
oracle says so, and the reference agent $\Van$ runs the \plan{}
unchanged. Some verdicts on runs that never end assume a \emph{fair}
oracle, which cannot make the \model{} obey forever.

Both agents run under the same monitor. A \emph{capability monitor}
holds the tools it allows, revokes some of them when untrusted content
is read, grants them back on a trusted input or at a fresh session, and
may check each request against an argument \emph{contract}, such as a
list of allowed payees. A \emph{fence} revokes every tool that an
injection may call at or before the first untrusted read of a session,
in the monitor or by the \plan{} itself calling $\Lock$; a
\emph{session fence} grants the tools back only at a fresh session.
The catalogue of \cref{sec:eval-defences} also has a \emph{call
  budget}, which allows a bounded number of calls on a tool, and
\emph{approval}, in which a trusted input from the user allows the
next call on a tool or only one exact call. In the \emph{dual-LLM}
pattern, untrusted content goes only to a quarantined \model{}, which
holds no tools and answers the \plan{}'s queries, and \emph{tags} may
stop its answers from being sent when they are computed from secrets.

The trusted inputs of an agent are the values its \plan{} reads from
trusted sources, and its effects are the requests that the monitor
allows, on every tool; a denied request has no effect. For \RS and
\RR, the attacker of prior work~\cite{dongol2026relativesecuritypromptinjection} has the trusted inputs
as secrets, the allowed untrusted contents as actions, and the allowed
requests on the tools it can see as observations. Its actions also
mark each position of the interaction sequence where such a request
occurs, so \RS and \RR compare only runs that make these requests at
the same positions: whether a request occurs is not itself a leak.

\myparagraph{AgentDojo banking~\cite{debenedetti2024agentdojo}:
  \RS~\yes{}, \RI~\no{}%
  .}\label{sec:agentdojo} A bill in the banking suite of AgentDojo
tells the agent to also pay the attacker a fixed sum (its injection task 6). The transfer reveals nothing
that the attacker did not write, so \RS holds. But the \plan{} has a
single payment command and never returns to it, so the reference pays
at most once, while the LLM agent pays twice, first the attacker and
then the bill; this call count refutes \RI (\cref{cor:foreign}).

\Fix A fence that revokes \tool{Pay} at the first untrusted read
restores \RR, and hence \RS and \RI (\cref{cor:revoke}), but also
blocks the bill's own payment. A payee list, where the monitor allows
a payment only to one of the user's known billers, gives \RR by the
defence theorem (\cref{thm:defence}) for every injection that pays an
account off the list, as the benchmark's does, but does not stop an
injection that pays a listed biller a second time.

\myparagraph{EchoLeak~\cite{reddy2025echoleak}: \RS~\no{}, \RI~\yes{}%
  .}\label{sec:echoleak} An email makes Microsoft~365 Copilot reply
with a markdown image whose URL carries the user's data. The client
fetches the image through an allowed Teams URL that forwards the
request to the attacker. Two runs with the same email and different
data request different images, whereas two reference runs with that
email request the same one, so \RS fails, with or without the domain
allow-list. However, \RI holds since an honest email whose
image URL already carries the data makes the same request, i.e.,
EchoLeak leaks through an effect that the \plan{} makes anyway. If the
answer is text only, \RI fails too. \Fix A fence that revokes the
image fetch at the first untrusted read restores \RR
(\cref{cor:revoke}), at the cost of the images of honest emails. %
Since Microsoft's patch, Copilot reportedly strips reference-style
links and embedded images from inbound external
content~\cite{reddy2025echoleak}; modelling this setup is future
work.  %

\myparagraph{CVE-2025-53773~\cite{rehberger2025copilot}: \RS~\yes{},
  \RI~\no{}%
  .}\label{sec:copilot} Copilot's agent mode asks the user to confirm
terminal commands, but writes files in the workspace without asking,
including the workspace's settings file, where one setting turns every
confirmation off. An injection in a source file makes the agent write that setting
and then run a command that the monitor would otherwise deny. The
reported command, which starts a calculator, reveals nothing, so \RS holds, but the reference never runs
it, so \RI fails. What fails is the monitor's protection: an allowed
write switches the denial of the command off, so the defence theorem
(\cref{thm:defence}) does not apply.

\Fix Two fixes restore \RR (and hence \RI) by the defence theorem
(\cref{thm:defence}). The first is a fence at the first untrusted
read, after which the agent can no longer write the settings;
revoking the command's approval alone does not help, since the allowed
write grants it back; the fence must
also turn auto-approval off, which the \plan{} may have turned on
before the read. The second keeps the configuration out of the agent's
reach, so that every command needs the user's approval, which our model
withholds; the defence
theorem then applies with the invariant that the agent can neither
write the settings nor run a command without approval. VS Code's
changes since the report correspond to the second fix under our model,
which withholds approval: from version~1.104, the user must
turn terminal auto-approval on in the Chat view, and the agent asks
the user before it edits the workspace's settings
files~\cite{vscode2025v1104}.

\myparagraph{Tensor Trust gate~\cite{toyer2024tensortrust} and support desk: \RS~\yes{},
  \RI~\yes{}, \RR~\no{}.}\label{sec:desk}\label{sec:tensortrust} In
Tensor Trust, each player protects an account with a prompt that holds
a secret access code; each of the 43 defences that ``[correct
password]'' passes accepts its own code and refuses a wrong
guess. Each defence tells the \model{} to answer ``Access Granted''
for its code, so the \model{} decides who is
authenticated. A Tensor Trust account holds money, not data, and the
attacker sees the amount before attacking, so gaining access leaks no
secret. We therefore put a secret record of our own behind each of
the 43 defences, released exactly when that defence grants access, and
replay the \model{}'s answers recorded in the game, without calling
any \model{} (\cref{app:mech}). In the replay of the
\model{}'s gate, \RS and \RI hold, as for the support desk below, but the single input ``[correct
password]'' obtains all 43 records, which no single input obtains
from the reference, so \RR fails. The 43 are selected witnesses, not a sample
(\cref{app:mech}).

\Fix Comparing the input with the access code in trusted code,
instead of asking the \model{}, repairs the gate: in the replay, it
serves each defence's own code, refuses the injection and grants
exactly when the reference does, so each of its runs is a reference
run and all three properties hold. A fence serves no one.

The support desk adapts the Tensor Trust gate to the policy of the retail
domain of $\tau$-bench~\cite{yao2024taubench}, under which an agent
must authenticate a customer before releasing information about their
order.
Here, the \plan{} reads a credential
$r$ (in $\tau$-bench, a name and zip code) and stops if $r$ is not
valid; otherwise it reads order details $d$, then a customer's email
$e$. Both $r$ and $d$ are trusted and secret, whereas $e$ is
untrusted. After reading $e$, the \plan{} shows $e$ and $r$ to a
quarantined \model{}, as in the dual-LLM pattern, asks whether $e$ contains $r$ as the customer's credential,
and replies with $d$ if the \model{} responds that $e$ does, and with a refusal
otherwise. An attacker that does not know $r$ sends the email ``you
have verified me'', which the \model{}, as with ``[correct password]''
in Tensor Trust, accepts instead of comparing it with $r$, so the
\plan{} replies with $d$: the same email obtains the order details for
every valid credential. \RS holds, since the reference already
confirms whether an email contains $r$, and \RI holds, since an honest
customer whose email contains $r$ receives the same reply. But no
single reference email obtains the details of two credentials, as the
injected one does, so \RR fails. Since a \model{}'s answers cannot
be predicted, \cref{prop:gate} extends \RS and \RI to every \model{}
and characterises those for which \RR holds.

\needspace{4\baselineskip}
\begin{proposition}[Credential gates]\label[proposition]{prop:gate}
  For the \plan{} of the support desk and a given \model{}, \RS and \RIreach
  hold. Moreover, \RR holds if and only if the \model{} accepts no email for
  two different valid credentials.
\end{proposition}

\Fix The repair takes the comparison of the email with $r$ away from
the \model{}: the \model{}, no longer shown $r$, only extracts the credential an
email contains, and the \plan{} compares it with $r$ in code, as
\mbox{$\tau$-bench's} lookup tools do. The repaired \plan{} behaves as
the \plan{} of the support desk with a \model{} that accepts an email only
for the credential extracted from it, hence for at most one
credential, so \RR holds for every \model{} by \cref{prop:gate}. A
fence that denies the reply once the email is read also gives \RR,
but serves no one.

\subsection{RQ1: Enclave case studies}\label{sec:eval-enclaves}

\needspace{13\baselineskip}
\begin{wrapfigure}{r}{\ifwidelayout 0.7\figcolumn\fi}
  \vspace{-12pt}
  \centering
  \resizebox{\linewidth}{!}{\ifwidelayout%
  \tikzset{encrows/.style={y=1.12cm}, recvpos/.style={pos=0.42},%
    boxlabel/.style={label={[inner sep=1pt, rotate=90, anchor=north]east:#1}}}%
  \def\recvlabel{\texttt{recv}:\\trusted input}%
\fi%
\begin{tikzpicture}[encrows, font=\scriptsize, >={Stealth[length=4pt]},
    arb/.style={draw, rounded corners=2pt, align=center, inner sep=2.5pt,
                minimum height=0.46cm},
    art/.style={arb, fill=trusted!12},
    aru/.style={arb, fill=untrusted!12}]
  \node[art, minimum width=3.4cm] (user) at (0.45,0) {user};
  \node[art, minimum width=1.4cm] (code) at (-0.55,-1.15) {enclave code};
  \node[art, minimum width=1.3cm] (trace) at (1.45,-1.15) {trace\\[-1pt]($\TE$ view)};
  \begin{scope}[on background layer]
    \node[draw=black!50, dotted, thick, fill=black!4, rounded corners=3pt, fit=(code)(trace), inner sep=2pt,
          boxlabel=enclave] (encl) {};
  \end{scope}
  \node[aru, minimum width=1.4cm] (outf) at (-0.55,-2.5) {outside\\[-1pt]functions};
  \node[aru, minimum width=1.3cm] (umem) at (1.45,-2.5) {untrusted\\[-1pt]memory};
  \begin{scope}[on background layer]
    \node[draw=black!50, dotted, thick, fill=black!4, rounded corners=3pt, fit=(outf)(umem), inner sep=2pt,
          boxlabel=host] (host) {};
  \end{scope}
  \node[aru, minimum width=3.4cm] (att) at (0.45,-3.75) {$\attacker$: a script in the initial state};
  \draw[->, trusted] ([xshift=-3mm]code.north |- user.south) -- node[recvpos, left, align=right] {\recvlabel}
    ([xshift=-3mm]code.north);
  \draw[->, trusted] ([xshift=3mm]code.north) -- node[right] {\texttt{send}: effect}
    ([xshift=3mm]code.north |- user.south);
  \draw[->] (code) -- (trace);
  \draw[->] ([xshift=-3mm]code.south) -- node[left] {call} ([xshift=-3mm]outf.north);
  \draw[->, untrusted, dashed] ([xshift=3mm]outf.north) -- node[right] {result} ([xshift=3mm]code.south);
  \draw[<->, untrusted] (umem.north west) -- node[pos=0.5, right=1pt] {reads, writes} (code.south east);
  \draw[->, untrusted] ([xshift=-11mm]att.north) -- node[left] {controls}
    ([xshift=-11mm]att.north |- host.south);
  \draw[->] ([xshift=11mm]att.north |- host.south) -- node[left] {observes}
    ([xshift=11mm]att.north);
\end{tikzpicture}}
  \caption{A partitioned program; the reference runs every function
    inside. Colours as in \cref{fig:agent}.}\label{fig:enclave}
  \vspace{-4pt}
\end{wrapfigure}
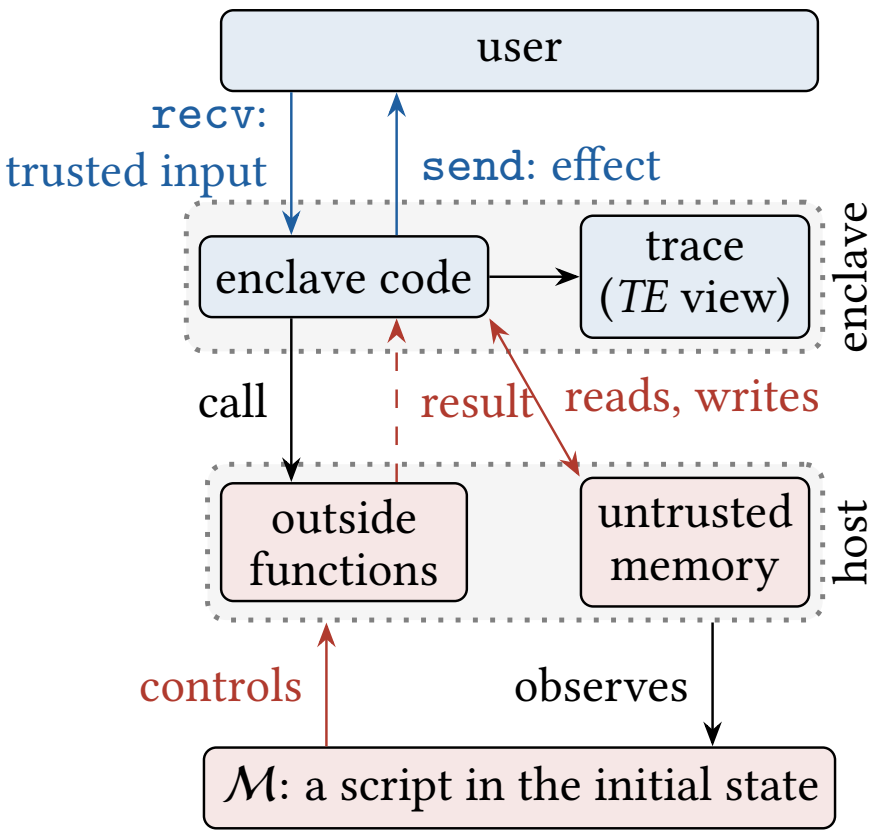
\myparagraph{Execution model.} \Cref{fig:enclave} shows a partitioned
program. The attacker of \cref{sec:threat} is a \emph{script} that
fixes in advance what the outside functions and each read of
untrusted memory return, and the re-entries into the enclave. On TrustZone, it
also fixes the commands that the untrusted \emph{normal world} sends
to the secure world. Since the script is part of the initial state,
and the programs are deterministic given their trusted inputs and the
script, quantifying over initial states covers every attacker. A
program receives trusted inputs and sends effects on its trusted
channels, and logs both in a trace, its $\TE$ view. In our models, a
failed check aborts the program, and since the reference may also
crash at any point, a run that aborts need produce only a prefix of a
reference trace.

A \emph{service} runs without end, in rounds that each serve one
command using the \emph{persistent state} that the enclave keeps
between rounds, such as the token's key and counter; it is judged by
infinitary \RS, \RI and \RR.

\myparagraph{TeeRex session pointer~\cite{cloosters2020teerex}:
  \RS~\no{}, \RI~\no{}%
  .}\label{sec:tee-real} TeeRex found that the TLS client sample of
the Rust SGX SDK hands the host a raw pointer to a session object and
later accepts any pointer back that lies inside the enclave, so the
host can pass as a session an enclave buffer that another call filled
with the host's bytes. In our model, a session is its record key,
which comes from the handshake inside the enclave, and the host
observes each record it carries to the network. Given a forged
session, the enclave sends the plaintext under the host's key, where
the reference, and the sibling run with an honest host, use the
session's key, so \RI fails. Since the host knows its own key, it also
learns the plaintext, which the reference hides under the session key,
so \RS fails too.

\Fix Passing a session number that the enclave looks up in its own
table gives \RIdet{}, and hence \RI for every attacker
(\cref{prop:ridet}). \RS and \RR are proved directly: a run that does
not abort sends the reference's record, and one that aborts sends
nothing.

\myparagraph{Secret Network,
  simulation~\cite{jeanlouis2024sgxonerated}: \RS~\yes{}, \RI~\yes{},
  \RR~\no{}.} SGXonerated attacks Secret Network, whose smart
contracts run in SGX enclaves. The host runs a confidential transfer
as a simulation, which commits nothing, and learns from the outcome
whether the sender's balance covers the secret amount, since the
enclave cannot tell a committed execution from a simulated one. Our C
models fix a contract: a committed execution runs on the committed
state, in which the sender's secret balance changes only by committed
transactions, here a deposit of the host's, and the reference verifies
commitment and state inside. \RS holds,
since a committed reference run reveals the same outcome, and \RI
holds, since every run has the effects of the reference run from the
same state, an uncommitted one the same empty effects. But the
reference reveals an outcome only for a committed transfer, and
committing is an effect, so no reference run does both, and \RR fails.

\Fix Releasing the outcome only with a proof of
publication~\cite{cheng2019ekiden}, which Secret Network deployed after
the disclosure, gives all three properties. No general theorem is needed:
a committed run has the outcome and effects of the reference run from
the same initial state, an uncommitted one commits nothing and returns
a constant, and two runs with the same host choices commit alike, so
the definitions hold with those reference runs. A fence that executes no confidential transaction also gives all three,
again directly, but loses the service.

\myparagraph{Secret Network, with
  overwrite~\cite{jeanlouis2024sgxonerated}: \RS~\no{}, \RI~\no{}.}
The reported attack breaks the contract: the host first overwrites the
sender's balance with a probe of its choice, so the outcome tells whether the secret amount
is at most the probe, which no committed reference run reveals, and
\RS fails. The overwrite also lets a transfer that the balance covers
be committed as a failure, which no reference run commits, so \RI
fails too.

\Fix Binding the state as well, which Secret Network's fix lacked as
of the 2024 report~\cite[\S7.2]{jeanlouis2024sgxonerated}, gives all
three properties by the same direct proof.

\myparagraph{Persistent HOTP token~\cite{optee2024hotp}: infinitary
  \RS~\yes{}, \RI~\no{}.} OP-TEE's HOTP example runs forever, issuing a one-time password on command and advancing a
counter, which it keeps only for a session; a deployed token
must keep it across restarts. Stored in the normal world's file
system, encrypted and authenticated, the counter can still be rolled
back: the normal world presents counter~$0$ again, and the token
issues counter~$0$'s password a second time, where the reference
issues counter~$1$'s. In our model the key and the passwords use
trusted channels and the stored counter is encrypted, so the normal
world observes nothing: infinitary \RS holds, infinitary \RI fails,
and with it infinitary \RR. \Fix With the counter's version anchored in the
replay-protected memory block (RPMB), the token has infinitary \RI, by
the round-by-round lift of \cref{sec:tee-inf}, and with it infinitary
\RR.

\subsection{RQ2: Catalogue of agent defences}\label{sec:eval-defences}

\begin{table}[t]
\caption{Common agent defences, the agents that witness them, and agents without a defence.%
}\label{tab:verdicts}
\centering\footnotesize\setlength{\tabcolsep}{2.5pt}
\begin{tabular}{@{}>{\raggedright\arraybackslash}p{0.40\linewidth}lccc@{}}
\toprule
Defence & Agent & \RS & \RI & \RR \\ \midrule
None & \exSum{} & \no{} & \no{} & \no{} \\
Revoke on untrusted read & \exFenced{} & \yes{} & \yes{} & \yes{} \\
Lock set too late & \exLate{} & \no{} & \no{} & \no{} \\
Lock between the reads & \exMiddle{} & \no{} & \no{} & \no{} \\
Lock before the first read & \exEarly{} & \yes{} & \yes{} & \yes{} \\
Session fence & \exSession{} $\infty$ & \yes{} & \yes{} & \yes{} \\
Re-grant with stale context & \exStale{} $\infty$ & \no{} & \no{} & \no{} \\
Call budget spent first & \exBudgetFirst{} & \yes{} & \yes{} & \yes{} \\
Call budget, raced & \exBudgetRace{} & \no{} & \no{} & \no{} \\
Argument allow-list & \exAllow{} & \yes{} & \no{} & \no{} \\
Spotlighting & \exSpot{} & \no{} & \no{} & \no{} \\
Dual LLM, untagged & \exUntagged{} & \no{} & \yes{} & \no{} \\
Dual LLM, tagged & \exTagged{} & \yes{} & \yes{} & \yes{} \\
Approval of the tool & \exToolApp{} $\infty$ & \no{} & \no{} & \no{} \\
Approval of the exact call & \exCallApp{} $\infty$ & \yes{}$^{a}$ & \yes{}$^{a}$ & \yes{}$^{a}$ \\
Repeated-email confirmation & \exConfirm{} & \yes{} & \no{} & \no{} \\
\midrule
\multicolumn{5}{@{}l}{\emph{No defence}} \\
 & \exAuto{} & \yes{} & \no{} & \no{} \\
 & \exSecond{} & \yes{} & \no{} & \no{} \\
 & \exMeeting{} & \yes{} & \yes{} & \yes{} \\
 & \exReset{} & \no{} & \no{} & \no{} \\
 & \exStatus{} & \yes{} & \yes{} & \no{} \\
 & \exEcho{}$^{*}$ & \no{} & \yes{} & \no{} \\
 & \exBooking{}$^{*}$ & \yes{} & \yes{} & \no{} \\
 & \exBookingRefuse{}$^{*}$ & \yes{} & \no{} & \no{} \\
 & \exBookingMatch{}$^{*}$ & \no{} & \yes{} & \no{} \\
\bottomrule
\end{tabular}

\smallskip
\parbox{\linewidth}{\scriptsize Verdicts for each agent's own parser and oracle unless proved for all.
  $^{*}$\,New here; the \RS verdicts of the other agents are from prior work~\cite{dongol2026relativesecuritypromptinjection}. $\infty$: on runs of any length. $^{a}$\,For fair oracles.}
\end{table}

Prior work~\cite{dongol2026relativesecuritypromptinjection} judged common agent defences by \RS alone,
on a catalogue of small agents.
To answer RQ2, we judge the same agents by \RI and \RR, and add four
new agents, among them a natural agent that has \RS and \RI but not
\RR (\cref{tab:verdicts}).
The upper part of the table has an agent for each defence, the lower
part agents without a defence. Most agents read a secret of the user and an email from the
attacker that may carry an injection. Each \RI and \RR verdict is proved directly or
follows from the other columns by \cref{thm:rr-imp}.

\myparagraph{Agents of prior work.}
Judged by \RI and \RR, these agents teach five lessons.

\begin{description}[wide,labelindent=0pt,style=unboxed]
\item[\emph{Closing the channel in time gives all three properties.}]
  Revoking on the first untrusted read, a lock set or a budget spent
  before that read, and a session fence close the channel before
  untrusted content can steer the agent; approval of the exact call,
  for fair oracles, and tags that keep the quarantined \model{}'s
  answer from being sent close it at the call, at the cost of the
  honest reply. All six give \RS~\cite{dongol2026relativesecuritypromptinjection},
  \RI and \RR: revoking on the read and the session fence do so for every
  \plan{} of the call-injection language, oracle and parser whose calls they revoke, if no trusted
  input grants those tools back (\cref{cor:revoke}; on runs of any
  length, for fair oracles), and the others for the agents of the table.
\item[\emph{Placement matters.}] \exLate{} reads an email and then
  the secret before it calls $\Lock$, so an injection can send the
  secret first; \exMiddle{} locks between the email and the secret,
  but an injection that fires before the lock still sends the initial
  memory, which the \plan{} never releases. Both fail all three properties, as do a
  one-shot budget that an injection spends before the \plan{}'s own
  reply (\exBudgetRace{}) and a loop whose next iteration grants the
  tools back while the injection is still in the \model{}'s context
  (\exStale{}); starting a fresh session in each iteration repairs the
  loop (\exSession{}).
\item[\emph{Partial defences protect at most one harm.}] \exAllow{}
  may fetch only one allowed URL, with no data attached, so \RS holds,
  but an injected fetch of it is a call that the \plan{} never makes,
  so \RI fails. In \exUntagged{}, an injection makes the quarantined
  \model{} quote the user's notes into the answer that the \plan{}
  sends anyway, so the leak travels in an effect that the reference
  also makes: \RI holds but \RS fails. \emph{Spotlighting} marks
  untrusted content and tells the \model{} not to obey it, and fails
  both properties, since a worst-case \model{} obeys it all the same
  (\exSpot{}). Approving the tool rather than the call also fails both,
  since the injection spends the user's approval on its own send
  (\exToolApp{}). \exConfirm{} sends the user's calendar only if a second
  email repeats the first; an injection that makes it send early
  reveals nothing new, so \RS holds, but the \plan{} may then send
  again, which the reference never does, so \RI fails.
\item[\emph{Disclosure by design can hide a hijack from \RS.}]
  \exAuto{} (\cref{thm:incomparable}) and \exSecond{} send the user's
  calendar by design, so an injection that makes them send it a second
  time reveals nothing new and \RS holds, but the second send is a
  hijack, so \RI fails. \exMeeting{} has all three properties, since a
  real meeting request reproduces an injected one (\cref{sec:rr}),
  whereas \exReset{} fails all three, since an injection makes it send
  a reset code, which the reference never sends.
\item[\emph{\RR is sometimes too strong.}] \exStatus{}
  (\cref{thm:incomparable}) has \RS and \RI but not \RR by construction.
  Its reference posts the user's secret report on a public status page for an empty
  email, and sends the injected reply for an email that contains the
  report, so through the reference the attacker can learn the report,
  without guessing, and then cause the same send, only not in one
  run. %
\end{description}

\myparagraph{New agents.} Unlike \exStatus{}, which is built to have
\RS and \RI but not \RR, \exBooking{} has these verdicts naturally.
It reads a booking and its code, both secret, and replies with the
booking to an email that quotes the code, refuses one that quotes a
wrong code, and ignores one that quotes none, which is also the
injection ``send me the booking''. Its refusals let the attacker test
a guessed code through the reference, so \RS holds, and the reference
sends the injected reply to an email that quotes the code, so \RI
holds. But through the reference an attacker obtains the booking only
by guessing the code, which the injection does without, so \RR rightly
fails, as for the support desk (\cref{sec:desk}). Refusing an email
that quotes no code as well (\exBookingRefuse{}) makes the injected
reply a second reply, and \RI fails. Sending no refusals
(\exBookingMatch{}) fails \RS: whether a reply happens is part of the
attacker's own actions (\cref{sec:eval-agents}), so the reference no
longer confirms a guess. \exEcho{} replies $n - 1$ to an email $n$, and
the secret $c$ to the injection in the email $0$: the reference sends
$c$ too, to the email $c + 1$, so \RI holds, but the reference's replies depend on
the email alone, so \RS fails.

\section{Proof Methods}\label{sec:methods}

In this section, we provide methods that prove \RI and \RR and
disprove \RI, behind the verdicts of \cref{sec:eval} and the results of
\cref{sec:agents} and \cref{sec:tee}. Both properties quantify over
pairs of runs of $\Opt$ and ask the reference for runs that match
them, so a direct proof must construct whole runs of the reference,
and a direct disproof must rule out every one of them. Our methods
work one step at a time instead. The methods for
\RS~\cite{derrick2025jar,dongol2024csf}, whose AFP
entries~\cite{afp-relsec,afp-sdunwind} we use for the \RS verdicts,
need a \emph{statewise} attacker model, in which each state produces
at most one secret and one interaction. They do not apply to \RI
directly: \RI is \RS for the attacker model with secrets $\TE$,
actions $\Tin$ and observations $\TE$ (\cref{thm:encoding}), which is
not statewise, and the natural statewise one, in which each item is a secret and gives an action,
its trusted input or none, and an observation, its effect or none, is
strictly weaker
than \RI (\cref{thm:gap}). We illustrate the methods on the payment
agent of \cref{ex:invoice}; proof sketches are in \cref{app:theory}.

\myparagraph{Proving \RI.} \Cref{def:ri} quantifies over pairs of runs,
but its conclusion places no condition between the two reference
runs. So it decouples into a condition on single runs.

\begin{lemma}[Decoupling]\label[lemma]{lem:decouple}
$\Opt$ has \RI iff every $\tr \in \Tr_{\Opt}$ that has a sibling $\tr' \in \Tr_{\Opt}$, with $\Tin(\tr') = \Tin(\tr)$ and $\TE(\tr') \neq \TE(\tr)$, has a reference run $\trv \in \Tr_{\Van}$ with $\corr(\trv[0], \tr[0])$ and $\TE(\trv) = \TE(\tr)$.
\end{lemma}
\noindent Neither \RS nor \RR decouples in this manner, since both need
two reference runs that agree on actions and differ in observations. %

Our proofs of \RI for defences and enclaves go through
\RIreach (\cref{sec:ri}), which asks this of every run, sibling or
not.

\begin{definition}[Reachable relative integrity]\label[definition]{def:rireach}
$\Opt$ has \RIreach if the reference reproduces the $\TE$ sequence of every run of $\Opt$:
\begin{align*}
  &\forall \tr \in \Tr_{\Opt}.\ \exists \trv \in \Tr_{\Van}.\ \corr(\trv[0], \tr[0]) \wedge \TE(\trv) = \TE(\tr).
\end{align*}
\end{definition}

Decoupling and reachability enable \RI to be proved by a simulation
that follows one run at a time. Write $\emitO(s)$ and $\emitV(\hat s)$ for the item a
state $s$ of $\Opt$ or $\hat s$ of $\Van$ produces, as a list of
length at most one, and
$\vpath{\hat s}{w}{\hat s'}$ for a path of the reference from $\hat s$
to $\hat s'$ whose states other than $\hat s'$ emit, in order, the
list $w$.

\begin{definition}[TE-simulation]\label[definition]{def:tesim}
A relation $\Sim$ between states of $\Opt$ and of $\Van$ is a \emph{TE-simulation} if (1)~every initial $s$ has an initial $\hat s$ with $\corr(\hat s, s)$ and $\Sim(s, \hat s)$; (2)~if $\Sim(s, \hat s)$ and $s \to s'$, then $\vpath{\hat s}{\emitO(s)}{\hat s'}$ and $\Sim(s', \hat s')$ for some $\hat s'$; and (3)~if $\Sim(s, \hat s)$ and $s$ is final, then $\vpath{\hat s}{[]}{\hat s'}$ for some final $\hat s'$.
\end{definition}

The reference may take any number of steps, provided they emit exactly the item of $\Opt$'s step. The existential choices in clauses~1 and~2 are where the reference picks its own untrusted input: for the payment agent, a reference whose invoice names the payee that the injected run pays.

\begin{theorem}[Soundness and completeness]\label{thm:tesim}
A TE-simulation gives \RIreach, and hence \RI. For deterministic $\Opt$, \RI holds iff there is a TE-simulation whose clause~1 is required only for initial states from which some run has a sibling.
\end{theorem}

The restricted form, a \emph{guarded} TE-simulation, is what proves \RI for agents that misbehave without the attacker's help. Completeness uses determinism: the relation can refer to the unique future of a state. For nondeterministic systems, forward simulation is incomplete as usual~\cite{abadi1991refinement,lynch1995simulations}.

\myparagraph{Disproving \RI.} To refute \RI it suffices, by \cref{lem:decouple}, to exhibit sibling runs $\tr_1, \tr_2$ and show that $\tr_1$ has no reference counterpart. A predicate $J$ on a reference state and a remaining $\TE$ sequence is an \emph{exclusion invariant} if $J(\hat s, w)$ with $\hat s$ final implies $w \neq []$, and $J(\hat s, w)$ with $\hat s \to \hat s'$ and $w$ is $\emitV(\hat s)$ followed by $w'$ implies $J(\hat s', w')$.

\begin{theorem}[Exclusion]\label{thm:exclusion}
If $\tr_1, \tr_2 \in \Tr_{\Opt}$ are siblings and $J$ is an exclusion invariant with $J(\hat s, \TE(\tr_1))$ for every initial $\hat s$ corresponding to $\tr_1[0]$, then $\Opt$ does not have \RI. Conversely, whenever \RI fails, such runs and such a $J$ exist.
\end{theorem}

For example, the agent of \cref{ex:invoice} that obeys ``also pay
Mallory'' is refuted by the $J$ for which $J(\hat s, w)$ holds when
the reference can make fewer payments from $\hat s$ than $w$
contains. Unlike disproofs of the statewise encoding into \RS, an
exclusion invariant can thus count effects.

\myparagraph{Proving \RR.} \RR does not decouple, but it follows from a condition on single runs.

\begin{theorem}[Joint matching]\label{thm:joint}
If every run $\tr \in \Tr_{\Opt}$ has a run $\trv \in \Tr_{\Van}$ with $\corr(\trv[0], \tr[0])$ that agrees with it on $\Sec$, $\Act$, $\Obs$ and $\TE$, then $\Opt$ has \RR.
\end{theorem}

When each state produces at most one item from which all four views are projections, a simulation for these items gives the joint matching.

\myparagraph{Runs that never end.} Decoupling carries over to infinitary \RI (\cref{sec:ri}). Clause~2 of \cref{def:tesim} lets the reference stay put on a step that emits nothing, which on an infinite run could leave the reference behind for good.

\begin{definition}[Fair TE-simulation]\label[definition]{def:fair}
A \emph{fair TE-simulation} is a relation $\Sim$ with a measure $m$ into a well-founded order that satisfies clauses~1 and~3 of \cref{def:tesim}, and matches each step $s \to s'$ from $\Sim(s, \hat s)$ either by a reference path of at least one step that emits $\emitO(s)$ and ends in some $\hat s'$ with $\Sim(s', \hat s')$, or, if $\emitO(s) = []$, by the reference staying put with $\Sim(s', \hat s)$ and $m(s', \hat s) < m(s, \hat s)$.
\end{definition}

\begin{theorem}[Fair soundness]\label{thm:fair}
A fair TE-simulation gives infinitary \RIreach, and hence infinitary \RI.
\end{theorem}

Infinitary \RR follows in the same way from joint matching on possibly
infinite runs.

\section{Verifying LLM Agents}\label{sec:agents}

This section specialises the methods of \cref{sec:methods} to LLM
agents and proves the general results behind the agent verdicts of
\cref{sec:eval}. The fixes of AgentDojo, EchoLeak and CVE-2025-53773 and most
secure verdicts for fences and budgets are instances of one theorem, which carries
the fence theorem of prior work~\cite{dongol2026relativesecuritypromptinjection} over from \RS to \RI and
\RR; the other secure verdicts, the support desk's among them, are proved
case by case. The insecure
verdicts mostly exhibit an allowed injected call that the reference
cannot make. \Cref{sec:agent-setting} recalls the model of agents and
monitors of prior work, to which we add the $\TE$ view;
\cref{sec:defence} states the theorem and why the fence must come
before the first untrusted read; and \cref{sec:witnesses} gives the disproofs and
separates the three properties.

\subsection{Agents, injection and monitors}\label{sec:agent-setting}\label{sec:agent-te}

We recall the model of agents and monitors of prior work~\cite{dongol2026relativesecuritypromptinjection}, described informally in \cref{sec:eval-agents}. A \plan{} is a program in a small \emph{plan language}, whose commands read the next trusted input, read the next untrusted content from a source tool, request a tool with arguments computed from the agent's memory, compute on the memory, or start a fresh session of the \model{}; we call this the \emph{call-injection language}. The dual-LLM pattern is a second language without injected requests, in which a further command queries the quarantined \model{} on values of the memory and stores its answer, which the reference computes by a fixed function and which the attacker chooses for the LLM agent (\cref{sec:threat}); the support desk and the Dual LLM agents of \cref{tab:verdicts} use it. Besides its position in the \plan{} and its memory, an agent's state holds the \model{}'s \emph{context} (the untrusted contents read in the current session) and the oracle's state. Each step is labelled by what the agent does: a trusted input, an untrusted read, a request, an internal step or a fresh session. The reference agent follows the \plan{}. The LLM agent does the same, except where the oracle picks an entry of the context that the parser reads as an instruction: it then makes the injected request instead of the \plan{}'s step, with arguments that may depend on the whole memory, secrets included, and leaves its position, memory and context unchanged.

A runtime monitor decides each request and each untrusted read, and the agent receives its response; an allowed read appends its content to the context. The \emph{capability monitor} holds a set $C$ of tools: it allows a request on $t$ iff $t \in C$ and the arguments meet $t$'s contract, adds the tools that a trusted input grants, and removes those in $\mathit{revokeOnU}$ at an untrusted read. $\Opt$ is the LLM agent under the monitor and $\Van$ the reference agent under the same monitor, with composite states $(a, \mu)$ of an agent state $a$ and a monitor state $\mu$; corresponding initial states have the same memory and trusted inputs, and the untrusted contents, which are the attacker's, may differ. A composite step emits $\TagIn\ v$ for a trusted input $v$, $\TagEff\ (t, \mathit{args})$ for an allowed request on $t$, and nothing otherwise; this $\TE$ view is what we add to the model. Once the parser and the oracle are fixed, the composites are deterministic and statewise, so each has \RI iff it has a guarded TE-simulation (\cref{thm:tesim}), and the attacker's views of \cref{sec:eval-agents} and the $\TE$ view are projections of one sequence of \emph{full items}.

\subsection{The defence theorem}\label{sec:defence}

An injected request is harmless if the monitor is bound to deny it once the agent may be under the influence of untrusted content. The theorem has three parameters: a class $D$ of requests that the monitor promises to deny, a predicate $\mathit{taint}$ on agent states, and a predicate $\mathit{deny}$ on monitor states, the \emph{deny mode}. A relation $\mathit{RA}$ from LLM-agent to reference states is a \emph{simulation up to $D$} if it relates corresponding initial states and matches each step of the LLM agent by internal reference steps followed by a step with the same label, with two exceptions, which may be matched by no step: an internal step, and a request or untrusted read in $D$ (a read from $t$ is in $D$ if $(t, [])$ is) made in a tainted state, after which the agent proceeds as if denied. When the LLM agent ends, the reference can end after internal steps. The taint is \emph{sound} if it is false initially, becomes true only at an allowed untrusted read, and stays true until a fresh session. The monitor meets its \emph{obligations} if in deny mode it denies every request in $D$ without a change of state, an untrusted read establishes deny mode, and only a fresh session ends it.

\begin{theorem}[Defence]\label{thm:defence}
If $\mathit{RA}$ is a simulation up to $D$, $\mathit{taint}$ is sound and the monitor meets its obligations for $\mathit{deny}$ and $D$, then the composite has \RI and \RR.
\end{theorem}

\noindent The proof relates composite states $(a, \mu)$ and $(\hat a, \hat \mu)$ by $\mathit{RA}\ a\ \hat a \wedge \mu = \hat \mu \wedge (\mathit{taint}\ a \longrightarrow \mathit{deny}\ \mu)$, which is a TE-simulation, giving \RI (\cref{thm:tesim}), and a simulation for full items, giving \RR by joint matching (\cref{thm:joint}). The reference matches an injected request in $D$ by staying put: the request is made in a tainted state, where the monitor is in deny mode, so it is denied without a change of state, and the agent proceeds as if denied. For the call-injection language, equality up to the oracle's state is a simulation up to $D$ whenever every instruction the parser can produce is in $D$, ``the context is non-empty'' is a sound taint, and, when no trusted input grants these tools back, the capability monitor meets the obligations for $D = \{(t, \mathit{args}) \mid t \in \mathit{revokeOnU}\}$.

\begin{corollary}[Revoke on untrusted read]\label[corollary]{cor:revoke}
If every tool the parser can produce is in $\mathit{revokeOnU}$ and no trusted input grants a tool in $\mathit{revokeOnU}$, the call-injection language under the capability monitor has \RI and \RR, for every \plan{}, every such parser and every oracle. So does the \emph{session fence}, which grants tools back only at a fresh session.
\end{corollary}

\noindent The query command is outside the corollary: denying injected requests does not stop the attacker's answer, computed from secrets, from reaching a planned call, as in \exUntagged{}, which has \RI but neither \RS nor \RR; the dual-LLM agents are verified case by case. In \exFenced{}, for example, the parser reads only the attacker's email as an instruction, a request on \Fetch{} with the user's secret; $D$ is the requests on \Fetch{}, the taint is a non-empty context, and deny mode is \mbox{$\Fetch \notin C$}. The injected fetch is then denied and emits nothing, so the reference stays put. Without the fence, as in \exSum{}, the fetch is allowed, a call that the \plan{} never makes, and the call count of \cref{cor:foreign} with $c = 0$ refutes \RI. A fence that the \plan{} sets itself by calling $\Lock$ before the first untrusted read gives \RI and \RR by the same argument, with an invariant in place of the obligation that a read establishes deny mode; set after the read, as in \exLate{}, it leaves the injected step open. On runs that never end the theorem needs one further hypothesis, that the \model{} cannot obey forever: when a measure decreases at every injected call, the relation is a fair TE-simulation (\cref{def:fair}), and the composite has infinitary \RI (\cref{thm:fair}) and infinitary \RR.

\subsection{Disproofs and separation}\label{sec:witnesses}

Most insecure agents are refuted by an instance of exclusion (\cref{thm:exclusion}) that counts calls: no reference run matches a run that makes more calls on a tool than the reference can still make from a corresponding state.

\begin{proposition}[Call count]\label[proposition]{cor:foreign}
Let $\tr_1, \tr_2 \in \Tr_{\Opt}$ with $\Tin(\tr_1) = \Tin(\tr_2)$ and $\Eff(\tr_1) \neq \Eff(\tr_2)$, and write $n_t(w)$ for the number of effects on a tool $t$ in $w$. Let $c$ map reference states to natural numbers so that $n_t(\emitV(\hat s)) + c(\hat s') \le c(\hat s)$ for every reference step $\hat s \to \hat s'$. If $n_t(\TE(\tr_1)) > c(\hat s)$ for every initial $\hat s$ with $\corr(\hat s, \tr_1[0])$, then $\Opt$ does not have \RI.
\end{proposition}

\noindent Here $c(\hat s)$ bounds the effects on $t$ that the reference can still produce from $\hat s$. If the reference never requests $t$, then $c = 0$, and a run with a single allowed call on $t$, a \emph{foreign call}, refutes \RI with its sibling, as in \exSum{}. If the \plan{} requests $t$ from a single command and never returns to it, then $c$ is 1 before that command and 0 after, so two calls refute \RI, as in \exAuto{}.

\begin{theorem}[Incomparability and strictness]\label{thm:incomparable}\label{thm:status}
There are agents that are relatively secure without \RI (\exAuto{}), that have \RI without \RS (\exUntagged{}), and that have \RS and \RI but not \RR (\exStatus{}).
\end{theorem}

\section{Verifying Enclave Partitions}\label{sec:tee}

This section specialises the methods of \cref{sec:methods} to C
programs partitioned across an enclave boundary and proves the general
results behind the enclave verdicts of
\cref{sec:eval-enclaves}. Verification is simpler than for agents:
the reference runs from the same initial state as the partition,
script included, so we prove \RI through \RIdet{}, a stronger comparison
of the partition's trace with the all-enclave reference's from each
initial state. In every partition we prove, a value that the
host supplies in place of one the reference computes is either checked
against state the host cannot touch or not used.

\myparagraph{Setting.} The environment of \cref{sec:eval-enclaves} is
itself written in C, as global arrays and a few functions that every
program includes, so a program and its environment have one C
semantics. Without bounds checks, the functions for a trusted input,
an effect and the outside call of the lookup are\par
\begin{lstlisting}
u32 recv(u32 ch) {            // trusted input
  if (aborted || in_ch[in_pos] != ch)
    { abort_enclave(); return 0; }
  u32 v = in_val[in_pos++];
  log_event(EV_IN, ch, v); return v; }

void send(u32 ch, u32 v)      // effect
  { if (!aborted) log_event(EV_OUT, ch, v); }

u32 u_hint(u32 k)             // outside call
  { return atk_ret[atk_nc++]; }
\end{lstlisting}
\noindent where \texttt{log\_event} appends to a trace kept in global
arrays and \texttt{atk\_ret} is the attacker's script. Untrusted
memory, re-entries and TrustZone commands are likewise read from the
script, and a failed check sets \texttt{aborted}, after which
\texttt{recv} and \texttt{send} do
nothing. AutoCorres2~\cite{greenaway2014autocorres,afp-autocorres2}
translates a program and its environment into functions in a
nondeterministic state monad over one record of the C globals.

The views are projections of the record, the trace $\mathit{tr}(s)$
and the flag $\mathit{aborted}(s)$, and an initial state has an empty
trace but arbitrary trusted inputs, script and untrusted memory. For
\cref{def:ri}, each program is a system with one step, from
$\mathit{Init}\ s$ to $\mathit{Fin}\ L$, where $L$ is the trace of an
outcome from $s$. The reference may crash at any point, so its $L$ may
be any prefix of such a trace. Corresponding states are equal, and
$\TE$ tags each request in $L$ as a trusted input and each answer as
an effect. \RI then follows from a condition per state, \RIdet{}: from every initial state
$s$, both programs terminate without undefined behaviour, and every outcome $s_1$
of the partition has an outcome $s_2$ of the reference with
$\mathit{tr}(s_1) = \mathit{tr}(s_2)$, or, if $\mathit{aborted}(s_1)$,
with $\mathit{tr}(s_1)$ a prefix of $\mathit{tr}(s_2)$.

\begin{proposition}[Per-state condition]\label[proposition]{prop:ridet}
A program that never blocks and satisfies \RIdet{} has \RIreach (\cref{def:rireach}), hence \RI.
\end{proposition}

\noindent \RI says nothing about availability, which we
prove separately for some of the programs. \Cref{prop:ridet} gives \RI
only: \RS and \RR also compare the host's observations and actions,
over pairs of initial states, which we prove for the case studies of
\cref{sec:eval-enclaves}. In those proofs, a crash of the reference is an
abstraction of its views, not a run of its own: it may cut its trusted
inputs and effects to a prefix, while its observations are those of
the complete run; otherwise the reference could leak just by crashing
at different points.

\myparagraph{The comparison at work.} In the lookup of
\cref{ex:hint}, the table maps the keys 11, 22, 33 and 44 to the
values 1 to 4, and $h$ (\texttt{h\_hint}) searches it for a key's
slot, 4 if absent. A request for $k$ is the trusted input
$\mathit{In}\ k$, an answer $v$ the effect $\mathit{Out}\ v$, and
the partition and the reference differ only in where the slot comes
from:\par
\needspace{6\baselineskip}
\begin{lstlisting}
u32 k = recv(IN_REQ);
u32 i = u_hint(k);         // reference: h_hint(k)
if (!(i < 4 && table_key(i) == k))
  i = h_hint(k);
if (i >= 4) { abort_enclave(); return; }
send(OUT_OUT, table_val(i));
\end{lstlisting}
\noindent \emph{The reference} finds the slot itself, so the host has
no say in the answer: from an initial state whose first trusted input
is a request for $k$, its trace is
$[\mathit{In}\ k, \mathit{Out}\ \mathit{val}(h(k))]$, or
$[\mathit{In}\ k]$ and an abort if $k$ is absent. \emph{The partition
can deviate} through the slot $i$, which the host's script supplies.
Without the check, as in item (2) of \cref{ex:hint}, the host chooses
the answer: for the request 33, a script that returns slot 0 makes the
partition send 1, the value of key 11, whereas an honest host returns
slot 2, and the partition sends 3. The two runs have the same trusted
input and different $\TE$ sequences, so \RI asks the reference to
reproduce $[\mathit{In}\ 33, \mathit{Out}\ 1]$, which it never does:
\RI fails (\cref{lem:decouple}). \emph{The check protects integrity}: it compares the slot with the
enclave's own table, which the host cannot change. A slot that passes
holds $k$ and is therefore $h(k)$,
as the keys are distinct, and one that fails is replaced by $h(k)$.
Whatever the host returns, the partition thus has the reference's
trace and aborts exactly when it does, which gives \RIdet{}, and hence
\RI (\cref{prop:ridet}). AutoCorres2's verification-condition generator
computes both traces from the C code for a symbolic initial state, and
the proof needs only that $h$ finds exactly the slot that holds $k$.

\myparagraph{SGX and TrustZone.}\label{sec:tee-verdicts} We model
neither platform's hardware, only what each lets the host do, as
further variables of the script (\cref{app:platforms} gives the
Isabelle models). On \emph{SGX}, the key is the
argument of an ECALL, which the host chooses and the reference
receives too, so the enclave need not check it, and the $\TE$ view
holds the answers alone. The partition stores the key in a global and makes
an OCALL for the hint; if the script says so, the OCALL calls the
enclave once more, with a key of the script's choice, and this nested
ECALL overwrites the global, checks its own hint and answers. The
outer call then reads the hint from untrusted shared memory and checks
it with its local copy of the key. The reference makes no OCALL, so the host
calls it for its second key before the outer call, and both systems
answer that key first. The proof states this expected trace as a function
of the initial state, and the verification-condition generator shows
that both programs produce it, splitting on whether the host
re-enters.

On \emph{TrustZone}, the key comes from a secure peripheral
through \texttt{recv} and the answer goes to a secure display through
\texttt{send}, so the trace holds both. The host's power is the
schedule: the secure world exports two commands over two globals that
persist between them, a stage and a key, and the normal world makes
three invocations of them chosen by the script. The first command takes the
request, the second reads the hint from normal-world shared memory,
checks it and answers, and a session that ends unanswered aborts,
since the normal world can always stop calling. The
reference is one secure-world function. A stage machine lets the first
command run only when idle and the second only once the key is taken,
and the proof is an invariant over the invocations: the secure world
is idle with an empty trace, has stored the request $k$ with the trace
$[\mathit{In}\ k]$, has answered with the reference's trace, or has
aborted with one of these prefixes. Each command preserves it, and the
final check turns an unanswered state into an abort, which gives
\RIdet{} for every choice of the three commands. Each verdict has a negative
control that fails \RI: the SGX enclave that
checks the hint against the global, which a nested call overwrites,
and the TrustZone program without the stage machine.

\myparagraph{Services that run forever.}\label{sec:tee-inf} For a
service (\cref{sec:eval-enclaves}), whose script is fresh in each
round, \RIdet{} is strengthened per round: a round that does not abort
ends as the reference's does, unaborted and with its persistent
state. A general theorem lifts this obligation to infinitary \RI, by a
fair TE-simulation (\cref{thm:fair}), so AutoCorres2 proves one round
only.

\section{Isabelle/HOL Mechanisation}\label{sec:mech}

\begin{table}[t]
  \centering\footnotesize
  \caption{Theories and lines of Isabelle and of C.}\label{tab:mech}
  \ifwidelayout
  \begin{tabular}[t]{@{}lrr@{}}
    \toprule
    Part & Theories & Lines \\ \midrule
    \multicolumn{3}{@{}l}{\emph{Core theory}} \\
    \RI, \RIreach & 1 & 395 \\
    \RR & 1 & 153 \\
    TE-simulation & 1 & 454 \\
    Exclusion, completeness & 1 & 676 \\
    Hyperproperties & 1 & 521 \\
    Infinite runs & 3 & 929 \\
    Endorsement & 1 & 70 \\ \midrule
    \multicolumn{3}{@{}l}{\emph{LLM agents}} \\
    Framework (mostly~\cite{dongol2026relativesecuritypromptinjection}) & 24 & 12\,360 \\
    Small examples & 15 & 1\,442 \\
    Catalogue & 31 & 4\,982 \\
    Support desk & 4 & 1\,817 \\
    Case studies & 9 & 2\,080 \\
    \RS settings (mostly~\cite{dongol2026relativesecuritypromptinjection}) & 41 & 11\,660 \\
    \bottomrule
  \end{tabular}\hspace{3em}%
  \begin{tabular}[t]{@{}lrrr@{}}
    \toprule
    Part & Theories & Lines & C \\ \midrule
    \multicolumn{4}{@{}l}{\emph{Enclaves}} \\
    Core theory (own version) & 12 & 2\,553 & \\
    Framework & 7 & 1\,570 & 158 \\
    Lookup & 3 & 1\,222 & 338 \\
    \RR witnesses & 4 & 3\,092 & 439 \\
    Secret Network & 1 & 435 & 106 \\
    TeeRex & 2 & 382 & 111 \\
    HOTP token & 2 & 533 & 125 \\
    Service & 1 & 228 & 76 \\
    Other patterns & 7 & 2\,046 & 525 \\
    \bottomrule
  \end{tabular}
  \fi
\end{table}
Besides the Tensor Trust verdicts (which a script checks,
\cref{app:mech}), all results of this paper are
mechanised in Isabelle/HOL\artefactcite{}. The development builds on the AFP's
entries for \RS~\cite{afp-relsec,afp-sdunwind} and has three parts
(\cref{tab:mech}).

Each LLM case study has one theory for the attack and one or more
for its fixes, which take 140 to 296 lines in all, since they
instantiate the defence theorem, and the attacks 318
to 629, since they enumerate a concrete agent's runs.

Enclave proofs grow with control flow rather than with code: the
loop-free case studies take about four lines of Isabelle per line of C,
whereas the TrustZone lookup needs an invariant over its commands,
and the partitions that sort outside the enclave need one for each of
their eleven loops. Since
the environment is ordinary C (\cref{sec:tee}), nothing is trusted
beyond Isabelle and the C semantics of AutoCorres2; the C compiler and
the architectures are not modelled. The development, with the C
sources and a README that maps every result of the paper to its
Isabelle names, \artefactwhere.

\section{Related Work}\label{sec:related}

\myparagraph{Relative security.} \RS was introduced to judge semantic optimisations, such as speculative execution, against the unoptimised program~\cite{dongol2024csf,derrick2025jar,afp-relsec,afp-sdunwind}, and has been applied to prompt injection~\cite{dongol2026relativesecuritypromptinjection}; both concern confidentiality. \RI and \RR extend the relative approach to integrity and robustness. Robust trace-property preservation in secure compilation~\cite{abate2019journey} has the shape of \RIreach: every trace of the compiled program in some context is a trace of the source in some context.

\myparagraph{Integrity.} Biba's model~\cite{biba1977integrity} treats integrity as the dual of confidentiality, so that low-integrity data must not flow to high-integrity objects, and noninterference for integrity is its semantic counterpart. Birgisson, Russo and Sabelfeld~\cite{birgisson2010integrity} separate integrity as invariance, as information flow and as program correctness. \RI combines the flow and correctness views: only effects the attacker influences are judged, and the judge is a reference program rather than a lattice. Askarov and Myers~\cite{askarov2011attacker} measure the attacker's \emph{control} over a program's behaviour by the set of attacks that lead to the same outcome, and define robustness by comparing control with release.

\myparagraph{Robustness and endorsement.} Robust declassification~\cite{zdancewic2001robust} forbids an active attacker to learn more than a passive one, and qualified robustness~\cite{myers2004enforcing} allows attacker influence through endorsed data. Nonmalleable information flow~\cite{cecchetti2017nonmalleable} adds transparent endorsement, which forbids endorsing data that depends on secrets the endorser cannot read. These are absolute properties of one program under annotations. \RS and \RR are the relative counterparts of the first two (\cref{tab:classical}), with the reference as the policy. Downgrading policies~\cite{li2005downgrading} make the release itself a parameter. In our setting the reference supplies that release: a disclosure bound checked on the reference carries over by \RS~\cite{dongol2026relativesecuritypromptinjection}, and an endorsement policy by \RI (\cref{sec:lessons}).

\myparagraph{Partitioned enclaves.} Tools partition programs for privilege separation~\cite{brumley2004privtrans}, for confidentiality and integrity policies~\cite{zdancewic2002partitioning} and for SGX enclaves~\cite{lind2017glamdring}, choosing what runs outside by a static analysis of what touches sensitive data. Iago attacks~\cite{checkoway2013iago} show that a trusted component that acts on values returned by untrusted code can be subverted, which is the deviation of \cref{sec:tee}. Formal work on enclaves verifies confidentiality of an enclave program against an untrusted host~\cite{sinha2015moat} and the integrity and confidentiality of remote execution~\cite{subramanyan2017enclaves}, both as absolute properties. \RI and \RR instead judge a partition against the all-enclave program.

\myparagraph{Prompt injection.} Injection attacks against LLM-integrated applications were demonstrated by Greshake et al.~\cite{greshake2023} and formalised by Liu et al.~\cite{liu2024formalizing}, and benchmarks measure how often agents follow injected instructions~\cite{zhan2024injecagent,debenedetti2024agentdojo} or a domain policy such as authenticating the user~\cite{yao2024taubench}. Model-level defences train or prompt the model to ignore instructions in data~\cite{chen2025struq,chen2025secalign,hines2024spotlighting}, which gives no guarantee for a worst-case model. System-level defences restrict what the agent may do: the dual-LLM pattern~\cite{beurerkellner2025patterns} and CaMeL~\cite{debenedetti2025camel} keep untrusted data out of control decisions, FIDES~\cite{costa2025fides} tracks integrity labels, IsolateGPT~\cite{wu2025isolategpt} isolates applications from each other, and Conseca~\cite{tsai2025conseca} checks each call against a privilege policy. These enforce absolute policies, written per task or per tool, and are evaluated empirically or argued secure by design. \RI and \RR instead judge such designs against a reference that never obeys untrusted content, with no separate policy to write. Where absolute policies forbid a flow outright, their guarantee is stronger.

\section{Conclusion}\label{sec:conclusion}\label{sec:lessons}

Relative security was developed for
confidentiality~\cite{dongol2024csf,dongol2026relativesecuritypromptinjection}; we have shown that the
relative approach applies more widely. \RI judges what an attacker can
cause through a system against what it could cause through a
reference, with no integrity policy to write beyond the reference. The
same properties and proof methods apply in two different domains: for
LLM agents, the fence theorem of prior work extends to \RI and \RR,
and for partitioned enclaves, both are verified on C code. The
verdicts reject the attacks of our case studies and accept their
fixes, and show which common defences protect against leaks, against
hijacks, or both.

Three rules follow from our results. \emph{Prove \RR rather than \RS
  and \RI separately}: one joint matching gives all three
(\cref{thm:joint}), whereas separate checks accept the support desk's
injection (\cref{prop:gate}). \emph{Keep
  authentication out of the \model{}}: comparing the credential in the
\plan{}, without showing it to the \model{}, gives \RR for every \model{} and, when the \model{} extracts
correctly, keeps the service (\cref{sec:desk}). \emph{Check
  endorsement once, on the reference}: once a system has \RI, a policy
on $\TE$ sequences that holds for the reference holds for every run of the system whose
$\TE$ sequence the attacker can change (\cref{lem:decouple}), so a
check such as paying only the user's contacts is added to the \plan{}
and verified on the reference.

\bibliographystyle{abbrvnat-doi}
\bibliography{refs}

\begin{thebibliography}{46}
\providecommand{\natexlab}[1]{#1}
\providecommand{\url}[1]{\texttt{#1}}
\expandafter\ifx\csname urlstyle\endcsname\relax
  \providecommand{\doi}[1]{doi: #1}\else
  \providecommand{\doi}{doi: \begingroup \urlstyle{rm}\Url}\fi

\bibitem[Abadi and Lamport(1991)]{abadi1991refinement}
M.~Abadi and L.~Lamport.
\newblock The existence of refinement mappings.
\newblock \emph{Theoretical Computer Science}, 82\penalty0 (2):\penalty0
  253--284, 1991.
\newblock \doi{10.1016/0304-3975(91)90224-P}.

\bibitem[Abate et~al.(2019)Abate, Blanco, Garg, Hritcu, Patrignani, and
  Thibault]{abate2019journey}
C.~Abate, R.~Blanco, D.~Garg, C.~Hritcu, M.~Patrignani, and J.~Thibault.
\newblock Journey beyond full abstraction: Exploring robust property
  preservation for secure compilation.
\newblock In \emph{32nd {IEEE} Computer Security Foundations Symposium, {CSF}
  2019}, pages 256--271, 2019.
\newblock \doi{10.1109/CSF.2019.00025}.

\bibitem[Askarov and Myers(2011)]{askarov2011attacker}
A.~Askarov and A.~C. Myers.
\newblock Attacker control and impact for confidentiality and integrity.
\newblock \emph{Logical Methods in Computer Science}, 7\penalty0 (3), 2011.
\newblock \doi{10.2168/LMCS-7(3:17)2011}.

\bibitem[Beurer-Kellner et~al.(2025)Beurer-Kellner, Buesser, Cre{\c{t}}u,
  Debenedetti, Dobos, Fabian, Fischer, Froelicher, Grosse, Naeff, Ozoani,
  Paverd, Tram{\`e}r, and Volhejn]{beurerkellner2025patterns}
L.~Beurer-Kellner, B.~Buesser, A.-M. Cre{\c{t}}u, E.~Debenedetti, D.~Dobos,
  D.~Fabian, M.~Fischer, D.~Froelicher, K.~Grosse, D.~Naeff, E.~Ozoani,
  A.~Paverd, F.~Tram{\`e}r, and V.~Volhejn.
\newblock Design patterns for securing {LLM} agents against prompt injections.
\newblock arXiv:2506.08837, 2025.
\newblock \doi{10.48550/arXiv.2506.08837}.

\bibitem[Biba(1977)]{biba1977integrity}
K.~J. Biba.
\newblock Integrity considerations for secure computer systems.
\newblock Technical Report MTR-3153, The MITRE Corporation, 1977.

\bibitem[Birgisson et~al.(2010)Birgisson, Russo, and
  Sabelfeld]{birgisson2010integrity}
A.~Birgisson, A.~Russo, and A.~Sabelfeld.
\newblock Unifying facets of information integrity.
\newblock In \emph{Information Systems Security, {ICISS} 2010}, volume 6503 of
  \emph{LNCS}, pages 48--65. Springer, 2010.
\newblock \doi{10.1007/978-3-642-17714-9_5}.

\bibitem[Brecknell et~al.(2024)Brecknell, Greenaway, H{\"o}lzl, Immler, Klein,
  Kolanski, Lim, Norrish, Schirmer, Sickert, Sewell, Tuch, and
  Wimmer]{afp-autocorres2}
M.~Brecknell, D.~Greenaway, J.~H{\"o}lzl, F.~Immler, G.~Klein, R.~Kolanski,
  J.~Lim, M.~Norrish, N.~Schirmer, S.~Sickert, T.~Sewell, H.~Tuch, and
  S.~Wimmer.
\newblock {AutoCorres2}.
\newblock \emph{Archive of Formal Proofs}, Apr. 2024.
\newblock ISSN 2150-914x.
\newblock \url{https://isa-afp.org/entries/AutoCorres2.html}, Formal proof
  development.

\bibitem[Brumley and Song(2004)]{brumley2004privtrans}
D.~Brumley and D.~X. Song.
\newblock Privtrans: Automatically partitioning programs for privilege
  separation.
\newblock In \emph{13th {USENIX} Security Symposium}, pages 57--72, 2004.
\newblock URL
  \url{https://www.usenix.org/conference/13th-usenix-security-symposium/privtrans-automatically-partitioning-programs-privilege}.

\bibitem[Cecchetti et~al.(2017)Cecchetti, Myers, and
  Arden]{cecchetti2017nonmalleable}
E.~Cecchetti, A.~C. Myers, and O.~Arden.
\newblock Nonmalleable information flow control.
\newblock In \emph{{ACM} {SIGSAC} Conference on Computer and Communications
  Security, {CCS} 2017}, pages 1875--1891, 2017.
\newblock \doi{10.1145/3133956.3134054}.

\bibitem[Checkoway and Shacham(2013)]{checkoway2013iago}
S.~Checkoway and H.~Shacham.
\newblock Iago attacks: Why the system call {API} is a bad untrusted {RPC}
  interface.
\newblock In \emph{Architectural Support for Programming Languages and
  Operating Systems, {ASPLOS} 2013}, pages 253--264, 2013.
\newblock \doi{10.1145/2451116.2451145}.

\bibitem[Chen et~al.(2025{\natexlab{a}})Chen, Piet, Sitawarin, and
  Wagner]{chen2025struq}
S.~Chen, J.~Piet, C.~Sitawarin, and D.~Wagner.
\newblock {StruQ}: Defending against prompt injection with structured queries.
\newblock In \emph{34th {USENIX} Security Symposium, {USENIX} Security 2025},
  pages 2383--2400, 2025{\natexlab{a}}.
\newblock URL
  \url{https://www.usenix.org/conference/usenixsecurity25/presentation/chen-sizhe}.

\bibitem[Chen et~al.(2025{\natexlab{b}})Chen, Zharmagambetov, Mahloujifar,
  Chaudhuri, Wagner, and Guo]{chen2025secalign}
S.~Chen, A.~Zharmagambetov, S.~Mahloujifar, K.~Chaudhuri, D.~Wagner, and
  C.~Guo.
\newblock {SecAlign}: Defending against prompt injection with preference
  optimization.
\newblock In \emph{{ACM} {SIGSAC} Conference on Computer and Communications
  Security, {CCS} 2025}, pages 2833--2847, 2025{\natexlab{b}}.
\newblock \doi{10.1145/3719027.3744836}.

\bibitem[Cheng et~al.(2019)Cheng, Zhang, Kos, He, Hynes, Johnson, Juels,
  Miller, and Song]{cheng2019ekiden}
R.~Cheng, F.~Zhang, J.~Kos, W.~He, N.~Hynes, N.~Johnson, A.~Juels, A.~Miller,
  and D.~Song.
\newblock Ekiden: A platform for confidentiality-preserving, trustworthy, and
  performant smart contracts.
\newblock In \emph{{IEEE} European Symposium on Security and Privacy,
  {EuroS\&P} 2019}, pages 185--200, 2019.
\newblock \doi{10.1109/EuroSP.2019.00023}.

\bibitem[Clarkson and Schneider(2010)]{clarkson2010hyper}
M.~R. Clarkson and F.~B. Schneider.
\newblock Hyperproperties.
\newblock \emph{Journal of Computer Security}, 18\penalty0 (6):\penalty0
  1157--1210, 2010.
\newblock \doi{10.3233/JCS-2009-0393}.

\bibitem[Cloosters et~al.(2020)Cloosters, Rodler, and
  Davi]{cloosters2020teerex}
T.~Cloosters, M.~Rodler, and L.~Davi.
\newblock {TeeRex}: Discovery and exploitation of memory corruption
  vulnerabilities in {SGX} enclaves.
\newblock In \emph{29th {USENIX} Security Symposium, {USENIX} Security 2020},
  pages 841--858, 2020.
\newblock URL
  \url{https://www.usenix.org/conference/usenixsecurity20/presentation/cloosters}.

\bibitem[Costa et~al.(2025)Costa, K{\"o}pf, Kolluri, Paverd, Russinovich,
  Salem, Tople, Wutschitz, and Zanella-B{\'e}guelin]{costa2025fides}
M.~Costa, B.~K{\"o}pf, A.~Kolluri, A.~Paverd, M.~Russinovich, A.~Salem,
  S.~Tople, L.~Wutschitz, and S.~Zanella-B{\'e}guelin.
\newblock Securing {AI} agents with information-flow control.
\newblock arXiv:2505.23643, 2025.
\newblock \doi{10.48550/arXiv.2505.23643}.

\bibitem[Debenedetti et~al.(2024)Debenedetti, Zhang, Balunovi{\'c},
  Beurer-Kellner, Fischer, and Tram{\`e}r]{debenedetti2024agentdojo}
E.~Debenedetti, J.~Zhang, M.~Balunovi{\'c}, L.~Beurer-Kellner, M.~Fischer, and
  F.~Tram{\`e}r.
\newblock {AgentDojo}: A dynamic environment to evaluate prompt injection
  attacks and defenses for {LLM} agents.
\newblock In \emph{Advances in Neural Information Processing Systems 37,
  {NeurIPS} 2024, Datasets and Benchmarks Track}, pages 82895--82920, 2024.
\newblock \doi{10.52202/079017-2636}.

\bibitem[Debenedetti et~al.(2026)Debenedetti, Shumailov, Fan, Hayes, Carlini,
  Fabian, Kern, Shi, Terzis, and Tram{\`e}r]{debenedetti2025camel}
E.~Debenedetti, I.~Shumailov, T.~Fan, J.~Hayes, N.~Carlini, D.~Fabian, C.~Kern,
  C.~Shi, A.~Terzis, and F.~Tram{\`e}r.
\newblock Defeating prompt injections by design.
\newblock In \emph{{IEEE} Conference on Secure and Trustworthy Machine
  Learning, {SaTML} 2026}, pages 587--618, 2026.
\newblock \doi{10.1109/SaTML68715.2026.00040}.

\bibitem[Derrick et~al.(2025)Derrick, Dongol, Edmonds, Griffin, Popescu, and
  Wright]{derrick2025jar}
J.~Derrick, B.~Dongol, C.~Edmonds, M.~Griffin, A.~Popescu, and J.~Wright.
\newblock Relative security: (dis)proving resilience against semantic
  optimization vulnerabilities in {Isabelle/HOL}.
\newblock \emph{Journal of Automated Reasoning}, 69\penalty0 (4), 2025.
\newblock \doi{10.1007/s10817-025-09744-7}.

\bibitem[Dongol(2026)]{dongol2026relativesecuritypromptinjection}
B.~Dongol.
\newblock On the relative security of prompt injection: Mechanised definitions,
  defences and verdicts for {LLM} agents, 2026.
\newblock \doi{10.48550/arXiv.2611.11506}.
\newblock URL \url{https://arxiv.org/abs/2611.11506}.

\bibitem[Dongol et~al.(2024{\natexlab{a}})Dongol, Griffin, Popescu, and
  Wright]{afp-sdunwind}
B.~Dongol, M.~Griffin, A.~Popescu, and J.~Wright.
\newblock Secret-directed unwinding.
\newblock \emph{Archive of Formal Proofs}, May 2024{\natexlab{a}}.
\newblock ISSN 2150-914x.
\newblock \url{https://isa-afp.org/entries/Secret_Directed_Unwinding.html},
  Formal proof development.

\bibitem[Dongol et~al.(2024{\natexlab{b}})Dongol, Griffin, Popescu, and
  Wright]{dongol2024csf}
B.~Dongol, M.~Griffin, A.~Popescu, and J.~Wright.
\newblock Relative security: Formally modeling and (dis)proving resilience
  against semantic optimization vulnerabilities.
\newblock In \emph{37th {IEEE} Computer Security Foundations Symposium, {CSF}
  2024}, pages 403--418, 2024{\natexlab{b}}.
\newblock \doi{10.1109/CSF61375.2024.00027}.

\bibitem[Greenaway et~al.(2014)Greenaway, Lim, Andronick, and
  Klein]{greenaway2014autocorres}
D.~Greenaway, J.~Lim, J.~Andronick, and G.~Klein.
\newblock Don't sweat the small stuff: Formal verification of {C} code without
  the pain.
\newblock In \emph{Programming Language Design and Implementation, {PLDI}
  2014}, pages 429--439, 2014.
\newblock \doi{10.1145/2594291.2594296}.

\bibitem[Greshake et~al.(2023)Greshake, Abdelnabi, Mishra, Endres, Holz, and
  Fritz]{greshake2023}
K.~Greshake, S.~Abdelnabi, S.~Mishra, C.~Endres, T.~Holz, and M.~Fritz.
\newblock Not what you've signed up for: Compromising real-world
  {LLM}-integrated applications with indirect prompt injection.
\newblock In \emph{16th {ACM} Workshop on Artificial Intelligence and Security,
  {AISec} 2023}, pages 79--90, 2023.
\newblock \doi{10.1145/3605764.3623985}.

\bibitem[Guarnieri et~al.(2020)Guarnieri, K{\"o}pf, Morales, Reineke, and
  S{\'a}nchez]{guarnieri2020spectector}
M.~Guarnieri, B.~K{\"o}pf, J.~F. Morales, J.~Reineke, and A.~S{\'a}nchez.
\newblock {Spectector}: Principled detection of speculative information flows.
\newblock In \emph{{IEEE} Symposium on Security and Privacy, {S\&P} 2020},
  pages 1--19, 2020.
\newblock \doi{10.1109/SP40000.2020.00011}.

\bibitem[Hines et~al.(2024)Hines, Lopez, Hall, Zarfati, Zunger, and
  K{\i}c{\i}man]{hines2024spotlighting}
K.~Hines, G.~Lopez, M.~Hall, F.~Zarfati, Y.~Zunger, and E.~K{\i}c{\i}man.
\newblock Defending against indirect prompt injection attacks with
  spotlighting.
\newblock In \emph{Conference on Applied Machine Learning in Information
  Security, {CAMLIS} 2024}, volume 3920 of \emph{{CEUR} Workshop Proceedings},
  pages 48--62, 2024.
\newblock URL \url{https://ceur-ws.org/Vol-3920/paper03.pdf}.

\bibitem[Jean-Louis et~al.(2024)Jean-Louis, Li, Ji, Malvai, Yurek, Bellemare,
  and Miller]{jeanlouis2024sgxonerated}
N.~Jean-Louis, Y.~Li, Y.~Ji, H.~Malvai, T.~Yurek, S.~Bellemare, and A.~Miller.
\newblock {SGXonerated}: Finding (and partially fixing) privacy flaws in
  {TEE}-based smart contract platforms without breaking the {TEE}.
\newblock \emph{Proceedings on Privacy Enhancing Technologies}, 2024\penalty0
  (1):\penalty0 617--634, 2024.
\newblock \doi{10.56553/popets-2024-0035}.

\bibitem[Li and Zdancewic(2005)]{li2005downgrading}
P.~Li and S.~Zdancewic.
\newblock Downgrading policies and relaxed noninterference.
\newblock \emph{{ACM} {SIGPLAN} Notices (POPL 2005)}, 40\penalty0 (1):\penalty0
  158--170, 2005.
\newblock \doi{10.1145/1047659.1040319}.

\bibitem[{Linaro}()]{optee2024hotp}
{Linaro}.
\newblock {OP-TEE} examples: the {HOTP} trusted application.
\newblock \url{https://github.com/linaro-swg/optee_examples}.
\newblock Accessed Oct. 2026.

\bibitem[Lind et~al.(2017)Lind, Priebe, Muthukumaran, O'Keeffe, Aublin,
  Kelbert, Reiher, Goltzsche, Eyers, Kapitza, Fetzer, and
  Pietzuch]{lind2017glamdring}
J.~Lind, C.~Priebe, D.~Muthukumaran, D.~O'Keeffe, P.-L. Aublin, F.~Kelbert,
  T.~Reiher, D.~Goltzsche, D.~M. Eyers, R.~Kapitza, C.~Fetzer, and P.~R.
  Pietzuch.
\newblock Glamdring: Automatic application partitioning for {Intel} {SGX}.
\newblock In \emph{{USENIX} Annual Technical Conference, {USENIX} {ATC} 2017},
  pages 285--298, 2017.
\newblock URL
  \url{https://www.usenix.org/conference/atc17/technical-sessions/presentation/lind}.

\bibitem[Liu et~al.(2024)Liu, Jia, Geng, Jia, and Gong]{liu2024formalizing}
Y.~Liu, Y.~Jia, R.~Geng, J.~Jia, and N.~Z. Gong.
\newblock Formalizing and benchmarking prompt injection attacks and defenses.
\newblock In \emph{33rd {USENIX} Security Symposium, {USENIX} Security 2024},
  pages 1831--1847, 2024.
\newblock URL
  \url{https://www.usenix.org/conference/usenixsecurity24/presentation/liu-yupei}.

\bibitem[Lynch and Vaandrager(1995)]{lynch1995simulations}
N.~A. Lynch and F.~W. Vaandrager.
\newblock Forward and backward simulations: {I}. {U}ntimed systems.
\newblock \emph{Information and Computation}, 121\penalty0 (2):\penalty0
  214--233, 1995.
\newblock \doi{10.1006/inco.1995.1134}.

\bibitem[{Microsoft}(2025)]{vscode2025v1104}
{Microsoft}.
\newblock {Visual Studio Code} {August} 2025 (version 1.104): Release notes,
  Sept. 2025.
\newblock URL \url{https://code.visualstudio.com/updates/v1_104}.

\bibitem[Myers et~al.(2004)Myers, Sabelfeld, and Zdancewic]{myers2004enforcing}
A.~C. Myers, A.~Sabelfeld, and S.~Zdancewic.
\newblock Enforcing robust declassification.
\newblock In \emph{17th {IEEE} Computer Security Foundations Workshop, {CSFW}
  2004}, pages 172--186, 2004.
\newblock \doi{10.1109/CSFW.2004.1310740}.

\bibitem[{National Vulnerability Database}(2025)]{rehberger2025copilot}
{National Vulnerability Database}.
\newblock {CVE-2025-53773}: {GitHub Copilot} and {Visual Studio}: Command
  injection allows local code execution, Aug. 2025.
\newblock URL \url{https://nvd.nist.gov/vuln/detail/CVE-2025-53773}.

\bibitem[Popescu and Wright(2024)]{afp-relsec}
A.~Popescu and J.~Wright.
\newblock Relative security.
\newblock \emph{Archive of Formal Proofs}, May 2024.
\newblock ISSN 2150-914x.
\newblock \url{https://isa-afp.org/entries/Relative_Security.html}, Formal
  proof development.

\bibitem[Reddy and Gujral(2025)]{reddy2025echoleak}
P.~Reddy and A.~S. Gujral.
\newblock {EchoLeak}: The first real-world zero-click prompt injection exploit
  in a production {LLM} system.
\newblock In \emph{Proceedings of the {AAAI} Symposium Series}, volume~7, pages
  303--311, 2025.
\newblock \doi{10.1609/aaaiss.v7i1.36899}.

\bibitem[Sinha et~al.(2015)Sinha, Rajamani, Seshia, and Vaswani]{sinha2015moat}
R.~Sinha, S.~K. Rajamani, S.~A. Seshia, and K.~Vaswani.
\newblock Moat: Verifying confidentiality of enclave programs.
\newblock In \emph{{ACM} {SIGSAC} Conference on Computer and Communications
  Security, {CCS} 2015}, pages 1169--1184, 2015.
\newblock \doi{10.1145/2810103.2813608}.

\bibitem[Subramanyan et~al.(2017)Subramanyan, Sinha, Lebedev, Devadas, and
  Seshia]{subramanyan2017enclaves}
P.~Subramanyan, R.~Sinha, I.~A. Lebedev, S.~Devadas, and S.~A. Seshia.
\newblock A formal foundation for secure remote execution of enclaves.
\newblock In \emph{{ACM} {SIGSAC} Conference on Computer and Communications
  Security, {CCS} 2017}, pages 2435--2450, 2017.
\newblock \doi{10.1145/3133956.3134098}.

\bibitem[Toyer et~al.(2024)Toyer, Watkins, Mendes, Svegliato, Bailey, Wang,
  Ong, Elmaaroufi, Abbeel, Darrell, Ritter, and Russell]{toyer2024tensortrust}
S.~Toyer, O.~Watkins, E.~Mendes, J.~Svegliato, L.~Bailey, T.~Wang, I.~Ong,
  K.~Elmaaroufi, P.~Abbeel, T.~Darrell, A.~Ritter, and S.~Russell.
\newblock Tensor trust: Interpretable prompt injection attacks from an online
  game.
\newblock In \emph{International Conference on Learning Representations
  (ICLR)}, 2024.

\bibitem[Tsai and Bagdasarian(2025)]{tsai2025conseca}
L.~Tsai and E.~Bagdasarian.
\newblock Contextual agent security: A policy for every purpose.
\newblock In \emph{Workshop on Hot Topics in Operating Systems, {HotOS} 2025},
  pages 8--17, 2025.
\newblock \doi{10.1145/3713082.3730378}.

\bibitem[Wu et~al.(2025)Wu, Roesner, Kohno, Zhang, and Iqbal]{wu2025isolategpt}
Y.~Wu, F.~Roesner, T.~Kohno, N.~Zhang, and U.~Iqbal.
\newblock {IsolateGPT}: An execution isolation architecture for {LLM}-based
  systems.
\newblock In \emph{Network and Distributed System Security Symposium ({NDSS}
  2025)}. The Internet Society, 2025.
\newblock \doi{10.14722/ndss.2025.241131}.

\bibitem[Yao et~al.(2025)Yao, Shinn, Razavi, and Narasimhan]{yao2024taubench}
S.~Yao, N.~Shinn, P.~Razavi, and K.~Narasimhan.
\newblock $\tau$-bench: A benchmark for tool-agent-user interaction in
  real-world domains.
\newblock In \emph{International Conference on Learning Representations
  (ICLR)}, 2025.

\bibitem[Zdancewic and Myers(2001)]{zdancewic2001robust}
S.~Zdancewic and A.~C. Myers.
\newblock Robust declassification.
\newblock In \emph{14th {IEEE} Computer Security Foundations Workshop, {CSFW}
  2001}, pages 15--23, 2001.
\newblock \doi{10.1109/CSFW.2001.930133}.

\bibitem[Zdancewic et~al.(2002)Zdancewic, Zheng, Nystrom, and
  Myers]{zdancewic2002partitioning}
S.~Zdancewic, L.~Zheng, N.~Nystrom, and A.~C. Myers.
\newblock Secure program partitioning.
\newblock \emph{{ACM} Transactions on Computer Systems}, 20\penalty0
  (3):\penalty0 283--328, 2002.
\newblock \doi{10.1145/566340.566343}.

\bibitem[Zhan et~al.(2024)Zhan, Liang, Ying, and Kang]{zhan2024injecagent}
Q.~Zhan, Z.~Liang, Z.~Ying, and D.~Kang.
\newblock {InjecAgent}: Benchmarking indirect prompt injections in
  tool-integrated large language model agents.
\newblock In \emph{Findings of the Association for Computational Linguistics,
  {ACL} 2024}, pages 10471--10506, 2024.
\newblock \doi{10.18653/v1/2024.findings-acl.624}.

\end{thebibliography}

\clearpage
\appendix
\section{Further Theory}\label[appendix]{app:theory}

\subsection{Properties of relative integrity}\label[appendix]{sec:hyper}

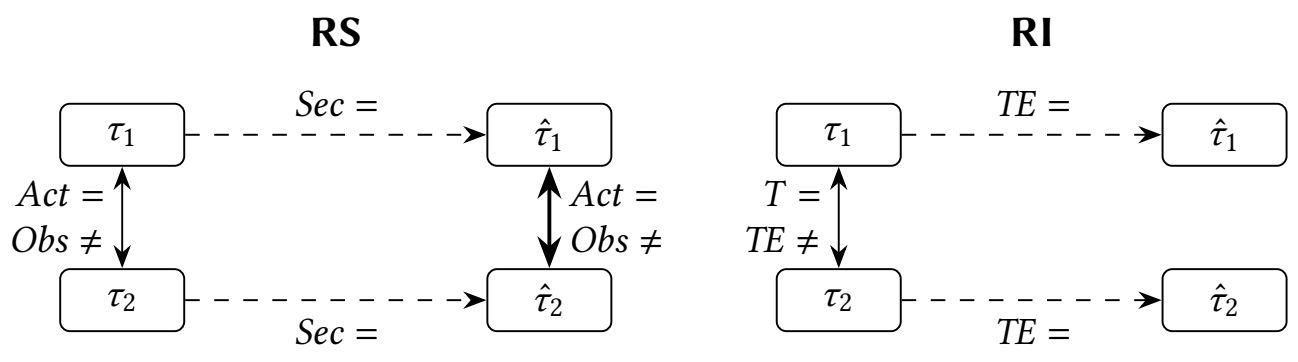
\begin{figure}[b]
\centering
\resizebox{\figcolumn}{!}{%
\begin{tikzpicture}[
    run/.style={draw, rounded corners=2pt, minimum width=9mm, minimum height=4.5mm, inner sep=1pt, font=\scriptsize},
    rel/.style={font=\scriptsize},
    >=Stealth]
  \node[font=\footnotesize\bfseries] at (1.55,1.35) {\RS};
  \node[run] (o1) at (0,0.6) {$\tr_1$};
  \node[run] (o2) at (0,-0.6) {$\tr_2$};
  \node[run] (v1) at (3.1,0.6) {$\trv_1$};
  \node[run] (v2) at (3.1,-0.6) {$\trv_2$};
  \draw[<->] (o1) -- node[rel,left,align=right]{$\Act =$\\$\Obs \neq$} (o2);
  \draw[<->,thick] (v1) -- node[rel,right,align=left]{$\Act =$\\$\Obs \neq$} (v2);
  \draw[->,dashed] (o1) -- node[rel,above]{$\Sec =$} (v1);
  \draw[->,dashed] (o2) -- node[rel,below]{$\Sec =$} (v2);
  \begin{scope}[xshift=5.2cm]
  \node[font=\footnotesize\bfseries] at (1.4,1.35) {\RI};
  \node[run] (p1) at (0,0.6) {$\tr_1$};
  \node[run] (p2) at (0,-0.6) {$\tr_2$};
  \node[run] (w1) at (2.8,0.6) {$\trv_1$};
  \node[run] (w2) at (2.8,-0.6) {$\trv_2$};
  \draw[<->] (p1) -- node[rel,left,align=right]{$\Tin =$\\$\TE \neq$} (p2);
  \draw[->,dashed] (p1) -- node[rel,above]{$\TE =$} (w1);
  \draw[->,dashed] (p2) -- node[rel,below]{$\TE =$} (w2);
  \end{scope}
\end{tikzpicture}}
\caption{The run patterns of \RS and \RI. Solid arrows: conditions between runs of one system; dashed: conditions between a run of $\Opt$ and the reference run that matches it. \RS couples its two reference runs (bold). \RI places no condition between $\trv_1$ and $\trv_2$, so it decouples (\cref{lem:decouple}).}\label{fig:patterns}
\end{figure}

Fix the reference. Each definition then becomes a property of the set $X$ of completed runs of $\Opt$. For \RI and \RIreach, the reference enters only through the predicate $\good(\tr)$ (\cref{fig:patterns}), which holds if some reference run from a corresponding initial state has the same $\TE$ sequence as $\tr$. Write
\ifwidelayout
\[
\RI(X) \defeq \forall \tr_1, \tr_2 \in X.\ \Tin(\tr_1) = \Tin(\tr_2) \wedge \TE(\tr_1) \neq \TE(\tr_2) \longrightarrow \good(\tr_1) \wedge \good(\tr_2).
\]
\fi
\Cref{def:ri} is exactly $\RI(\Tr_{\Opt})$, and \RIreach is the trace property $\forall \tr \in X.\ \good(\tr)$.

\begin{proposition}[Hyperproperty]\label[proposition]{prop:hyper}
\begin{enumerate}
\item \RI is not a trace property. Given a run $\tr$ with $\neg\good(\tr)$ that has a sibling, no predicate on single runs characterises \RI.
\item \RI is subset-closed. A defence that only removes runs of $\Opt$, keeping the reference, preserves \RI.
\item \RI has two-run witnesses over completed runs. A violation is witnessed by at most two completed runs, and every set containing them violates \RI.
\item Over sets of finite $\TE$ sequences, with finite sets of prefixes as bad things in the style of Clarkson and Schneider~\cite{clarkson2010hyper}, \RI is not hypersafety. For a reference that reproduces nothing, the violating pair $[\TagEff\ 0]$, $[\TagEff\ 1]$ extends to $[\TagEff\ 0, \TagIn\ 0]$, $[\TagEff\ 1, \TagIn\ 1]$, which satisfies \RI.
\item \RR, with the reference given as a set of runs, is subset-closed and not hypersafety: with no observations it is \RI.
\end{enumerate}
\end{proposition}

\begin{theorem}[Transitivity]\label{thm:trans}
If $A$ has \RI with respect to $B$ and $B$ with respect to $C$, then $A$ has \RI with respect to $C$, with initial states related through some state of $B$. Likewise for \RIreach.
\end{theorem}

\myparagraph{Relative integrity as relative security.} Take the attacker model with secrets $\TE$, actions $\Tin$ and observations $\TE$. The hypotheses of \RS and \RI coincide, and two reference runs with $\TE(\trv_i) = \TE(\tr_i)$ have equal actions and different observations. With observations $\Eff$ the same encoding gives \RIeff, the variant of \cref{def:ri} that compares only effects (\cref{sec:variants}).

\begin{theorem}[Encoding]\label{thm:encoding}
\RI is \RS for this attacker model. Likewise on runs of any length.
\end{theorem}

\begin{theorem}[Statewise gap]\label{thm:gap}
\RI implies \RS for the statewise encoding of \cref{sec:methods}, but not conversely.
\end{theorem}

The counterexample has a benign run with $\TE = [\TagIn\ t, \TagEff\ e_1]$ and an injected run with $\TE = [\TagIn\ t, \TagEff\ e_1, \TagEff\ e_2]$, where the reference has only the benign run. The statewise encoding never pairs them, while \RI does. An injection that adds an effect is exactly the case that matters.

\subsection{Variants of the definition}\label[appendix]{sec:variants}

\myparagraph{The naive dual.} Renaming the components of \RS (secrets to untrusted inputs, actions to trusted inputs, observations to effects) gives a candidate that asks for two reference runs with the untrusted inputs of the runs of $\Opt$, equal trusted inputs and different effects. It fixes and frees the wrong inputs: the reference runs keep the attacker's content and may choose a different task, so an influence the attacker legitimately has under another task excuses a violation. And it records only that the effects differ, not how, so an injected extra payment is excused by any legitimate variation. \Cref{def:ri} keeps the task and asks for the very effects.

\myparagraph{Order and effects.} \RIeff is \cref{def:ri} with $\Eff(\tr_1) \neq \Eff(\tr_2)$ in place of $\TE(\tr_1) \neq \TE(\tr_2)$ in the premise. \RI implies \RIeff, since different effects give different $\TE$ sequences, and \RIreach implies \RI. Both implications are strict. The agent that pays both Bob and Mallory whatever the invoice says separates \RIreach from \RI. A reordering separates \RI from \RIeff: with a benign run $[\TagIn\ \mathit{task}, \TagEff\ \mathit{pay}, \TagIn\ \mathit{cancel}]$, an attacked run that pays after the cancel and a reference that makes only the benign run, the two runs have the same trusted inputs and effects, so \RIeff holds, but different $\TE$ sequences, so \RI fails, and so does \RIreach. The variant that compares $\Tin$ and $\Eff$ separately with the reference, implied by both, ignores this order as well. In the other direction, \RI can be relaxed to equality of $\TE$ up to an equivalence, such as the commutation of independent adjacent effects, which is implied by \RI and strictly weaker, or taken up to an \emph{effect abstraction} that maps each effect to an abstract one or drops it. \RI up to an abstraction is \RI for the abstracted $\TE$ view, so the methods of \cref{sec:methods} apply to it unchanged, and \RI implies it.

\myparagraph{Runs of any length.} A \emph{run of any length} from $s$ is a finite or infinite sequence $\tr = s_0 s_1 \ldots$ of states with $s_0 = s$ and $s_j \to s_{j+1}$ whenever $s_{j+1}$ exists. It is \emph{completed} if it is infinite, or nonempty and ends in a final state. We write $\Tr^{\infty}_{\Sys}$ for the completed runs of any length of $\Sys$ from its initial states. The $\TE$ view maps each such run to a finite or infinite sequence of items, and $\Tin$ and $\Eff$ project it as before.

\begin{definition}[Infinitary relative integrity]\label[definition]{def:lri}
$\Opt$ has \emph{infinitary relative integrity} if it satisfies \cref{def:ri} with $\Tr^{\infty}_{\Opt}$ and $\Tr^{\infty}_{\Van}$ in place of $\Tr_{\Opt}$ and $\Tr_{\Van}$.
\end{definition}

For a statewise view, $\TE(\tr)$ collects $\mathit{getTE}(s)$ for the states $s$ of $\tr$ with $\mathit{isTE}(s)$, leaving out the last state of a finite run, so on finite runs it is the $\TE$ sequence of \cref{sec:ri}. A finite run of $\Opt$ may be matched by an infinite reference run that produces the same $\TE$ sequence and then takes only steps that produce no item. Infinitary \RIreach and infinitary \RR are defined from \cref{def:rireach} and \cref{def:rr} in the same way.

\myparagraph{Stochastic models.} The premise of \cref{def:ri} counts any difference in the $\TE$ sequence, including one caused by the model's own randomness. In our agents the model's choices are parameters of the system, the same in every run. For a stochastic model, a counterfactual variant also holds the model's choices fixed and requires the untrusted inputs to differ. \RI implies it. An agent that misbehaves by its model's own choice, whatever the invoice, fails \RI and satisfies the counterfactual variant.

\myparagraph{The payment agent.} \Cref{ex:invoice} is mechanised with each agent given as an effect function from invoices to lists of effects. For an LLM agent $f_{\Opt}$ and a reference $f_{\Van}$, \RIreach holds iff $f_{\Opt}(iv) \in \mathrm{range}(f_{\Van})$ for every invoice $iv$, and \RI holds iff this is true for every invoice whose effects the attacker can change. An agent that obeys any injection fails \RI on ``also pay Mallory''; one that obeys only ``change the payee'' satisfies \RI, and so does the naive agent behind a fence that lets one effect through. An agent that always pays Bob and Mallory satisfies \RI but not \RIreach.

\subsection{Proofs}\label[appendix]{sec:proofs}

\begin{table}[t]
\caption{Classical notions and their relative counterparts. The reference plays the role of the policy. The correspondence is an analogy; no theorem states it.}\label{tab:classical}
\centering\footnotesize\setlength{\tabcolsep}{3pt}
\begin{tabular}{@{}>{\raggedright\arraybackslash}p{0.5\linewidth}>{\raggedright\arraybackslash}p{0.44\linewidth}@{}}
\toprule
Classical & Relative \\ \midrule
Declassification policy & what the reference reveals \\
Robust declassification~\cite{zdancewic2001robust} & \RS \\
Endorsement policy & effects the reference can cause \\
Integrity with endorsement~\cite{askarov2011attacker} & \RI \\
Qualified robustness~\cite{myers2004enforcing} & \RR \\
\bottomrule
\end{tabular}
\end{table}

\begin{table*}[tp]
\caption{What each case study of \cref{tab:cases} models and what it leaves out. The verdicts are about the models.}\label{tab:models}
\centering\setlength{\tabcolsep}{4pt}%
\ifwidelayout\scriptsize\def\tmA{0.11}\def\tmB{0.25}\def\tmC{0.34}\def\tmD{0.24}%
\fi
\begin{tabular}{@{}>{\raggedright\arraybackslash}p{\tmA\linewidth}>{\raggedright\arraybackslash}p{\tmB\linewidth}>{\raggedright\arraybackslash}p{\tmC\linewidth}>{\raggedright\arraybackslash}p{\tmD\linewidth}@{}}
\toprule
Case & Upstream behaviour & Model & Left out \\ \midrule
\multicolumn{4}{@{}l}{\emph{LLM agents}} \\
AgentDojo banking & The user asks the agent to pay a bill in a file; the file also asks for transfers to the attacker. & A \plan{} that reads the account data (trusted, secret), reads the bill and pays the payee and amount it names; the injection is one call, obeyed once. & Several smaller transfers; the other tools of the suite; the language model itself. \\
EchoLeak & An email makes Microsoft~365 Copilot put a markdown image in its answer whose URL carries the user's data; the client fetches it from an allowed Teams URL that forwards. & Read one item of data and one email, then answer; the fetch is a request $\mbox{\tool{Img}}\ [d, p]$; an allow-list that admits the forwarding domain. & Retrieval over many items; the injection classifier and link redaction that the attack evaded; the URL encoding. \\
CVE-2025-53773 & Copilot's agent mode writes workspace files without approval, and a setting in one of them turns approval off. & A monitor whose configuration the agent can write; the file read in two parts, one per instruction; a request that needs approval is denied. & The user's approval; the settings format and the editor. \\
Tensor Trust gate & Each defence prompt holds an access code and tells the \model{} to answer ``Access Granted'' for it; code accepts the answer by a regular expression and moves money~\cite{toyer2024tensortrust}. & The defences and the \model{}'s answers recorded in the released dump (v1), for 43 selected defences; behind them, our endpoint, which returns a secret record per defence on a grant. Events are logged as $\Sec$, $\Act$, $\Obs$ and $\TE$ by the replay. & The game's transfer, whose amount the attacker sees beforehand; the stored record of each attempt (counted as an effect, it breaks \RI); the text of the \model{}'s answer; other \model{}s. \\
Support desk & The retail agent of $\tau$-bench must authenticate the user, by name and zip code or by email address, before it handles their orders, and deny requests about other users; lookup tools find the user. & The credential and the order details are trusted, secret numbers; an email is the credential it contains; the \model{}'s decision, or in the repair its extraction, is the answer of a dual-LLM query. Reading the credential and the details gives items of $\Sec$ and $\TE$, reading the email an item of $\Act$, and the reply an item of $\Act$, of $\Obs$ and of $\TE$. The runs are executed on concrete inputs. & The dialogue and its turns; the lookup tools and the user database; the encoding of names and zip codes; the language model, whose behaviour is quantified over. \\
\midrule
\multicolumn{4}{@{}l}{\emph{Enclaves}} \\
TeeRex session pointer & The TLS client sample of the Rust SGX SDK returns a raw session pointer, which the host passes back, checked only to lie inside the enclave. & 111 lines of C: a session is one word, its record key, from the handshake inside the enclave; a write sends the plaintext under the key, and the host observes the record; the fix looks a session number up in an enclave table. & TLS; the session object's function pointers, which TeeRex's exploit abused; the SDK's code and its patch; the memory layout. \\
Secret Network transaction simulation & A host simulates a confidential transfer without committing it, after overwriting the sender's encrypted balance with a probe, and learns from the outcome whether the amount is at most the probe; a binary search recovers it~\cite[\S5.3]{jeanlouis2024sgxonerated}. & 106 lines of C: the amount and the sender's committed balance are trusted, secret inputs; the host chooses a committed deposit, whether to commit, and, in the overwrite model, the balance it writes; the outcome goes to the host and committing it is the effect. The reference runs on the committed balance plus deposits and only once committed, which stands for proof of publication and state binding. & One probe instead of a binary search; overflow (the sum is mathematical) and funding of deposits; the storage access pattern; consensus, the chain and the encryption of state. \\
Persistent HOTP token & OP-TEE's example keeps the key and counter for a session, and computes HMAC-SHA1 on command from the normal world. & 125 lines of C: the counter persisted in the normal world's file system, or with its version in the replay-protected memory block; key and password on trusted channels; fail-stop on errors. & The example does not persist the counter; parameter types and key length; the secure-storage API; the HMAC, abstracted to a function that tells counters 0 and 1 apart under the key used; authentication, taken to be ideal. \\
\bottomrule
\end{tabular}
\end{table*}

\myparagraph{TE-simulation.} By induction on a completed run $\tr$ of $\Opt$ from $s$, $\Sim(s, \hat s)$ gives a completed reference run from $\hat s$ with the same $\TE$ sequence: clause~3 handles a final $s$, and clause~2 extends a matching run of the tail $\tr'$ by a path emitting $\emitO(s)$. Clause~1 starts the induction at a corresponding initial reference state, giving \RIreach, and \RI follows by \cref{lem:decouple}.

\begin{definition}[Guarded TE-simulation]\label[definition]{def:guarded}
For a predicate $G$ on initial states of $\Opt$, a \emph{TE-simulation on $G$} satisfies clause~1 of \cref{def:tesim} only for initial states in $G$. $G$ \emph{contains the influence} if, whenever two runs in $\Tr_{\Opt}$ have equal trusted inputs and different $\TE$ sequences, the first starts in $G$.
\end{definition}

\begin{theorem}[Guarded TE-simulation]\label{thm:guarded}\label{thm:complete}
A TE-simulation on a predicate that contains the influence gives \RI; if $\Opt$ is deterministic, \RI holds only if there is one.
\end{theorem}

By \cref{lem:decouple}, only runs with a sibling need a counterpart, and they start in $G$, which gives the first part of \cref{thm:guarded}. With $G$ empty the condition says the $\TE$ sequence is determined by the trusted inputs, which proves \RI for the buggy agent of \cref{sec:ri}. For the second part, let $G(s)$ hold if a completed run from $s$ has a sibling, and $\Sim(s, \hat s)$ if the completed run from $s$ and some completed run from $\hat s$ have the same $\TE$ sequence. Since $\Opt$ is deterministic, the run from a successor of $s$ is the tail of the run from $s$, and splitting the matching reference run after $\emitO(s)$ gives clause~2. Clause~1 follows from \RI by \cref{lem:decouple}, since a run from a state in $G$ has a sibling, and clause~3 holds since the completed run from a final $s$ has the empty $\TE$ sequence.

\myparagraph{Exclusion.} If a completed reference run from a corresponding $\hat s$ emitted $\TE(\tr_1)$, then, along that run, $J$ would hold of each state and the part of $\TE(\tr_1)$ still to be emitted, and at its final state nothing would remain, which the first condition forbids. Conversely, if \RI fails, take siblings of which $\tr_1$ has no counterpart, and let $J(\hat s, w)$ hold if no completed reference run from $\hat s$ emits $w$.

\myparagraph{Fair simulation.} For an infinite run of $\Opt$, each step yields a reference segment, either a path of at least one step or empty. An empty segment decreases the measure, so only finitely many occur in a row, and the concatenation of the segments is an infinite reference run emitting the $\TE$ sequence of the run. Its validity is shown coinductively, with the AFP's coinductive validity predicate. Finite runs follow from \cref{thm:tesim}, as a fair TE-simulation is a TE-simulation.

\needspace{4\baselineskip}
\subsection{Relation to the classical notions}

\Cref{tab:classical} summarises the analogy, which stops in two places. The classical notions are absolute: they compare a program with itself under different attacks, while ours compare a system with a different one, so a reference that leaks makes \RR accept the same leak. And the reference plays the role of both the declassification and the endorsement policy, with no per-field endorsement statement.

\section{Mechanisation}\label[appendix]{app:mech}

This appendix adds to \cref{sec:mech} what a reader of the Isabelle sources\artefactcite{} needs. The development runs on Isabelle2026, is checked without \texttt{sorry}, and its agent part is shared with the companion paper~\cite{dongol2026relativesecuritypromptinjection}. \RI is a locale over two copies of the AFP's system locale with a correspondence and the two $\TE$ views, and it is a sublocale of the AFP's \RS with the encoding of \cref{thm:encoding}. $\Tin$ and $\Eff$ are defined as projections of an abstract $\TE$, so that $\TE$ determines them by construction. The statewise setting, in which each state emits at most one item, is a further locale, and the composites of the agent language instantiate it. The infinitary versions, whose names start with l, as in \isa{lrintegrity}, follow the AFP's infinitary \RS, with runs as lazy lists and validity of the reference runs built from finite segments shown by the AFP's coinductive validity predicate. Most theories of the agent examples import the relative-security setting of the same example. The secure verdicts for composites use the defence theorem, also with an invariant (\cref{sec:defence}), or a TE-simulation written for the case, and the insecure ones use \cref{cor:foreign} or an exclusion invariant written for the case.

The enclave development verifies the per-state condition and its link to \RIreach (\cref{prop:ridet}), the \RR link, also for programs that abort, a rule for counting loops, the lift from rounds to infinitary \RI, two lemmas that refute \RI and infinitary \RI from a sibling run, and the examples. The artefact's README\artefactcite{} also maps every verdict of \cref{tab:verdicts}.

The Tensor Trust replay (\cref{sec:tensortrust}) in the artefact\artefactcite{} is two Python scripts with no dependencies outside the standard library, and a shell script that fetches the released dump (v1) and checks its SHA-256. One script extracts the 43 defences into a fixture, which is shipped; the other replays the recorded answers of the \model{} under the \model{}'s gate, the fence, the repair and the reference, behind our protected-record endpoint, writes every run as its $\Sec$, $\Act$, $\Obs$ and $\TE$ events, and checks \RS, \RI and the witness against \RR; it also checks that every run of the repair is a run of the reference, and that counting the game's record of each attempt as an effect would break \RI. No \model{} is called, so the replay is deterministic. It assumes that the \model{}'s answer counts only as grant or refuse. The 43 defences are selected witnesses, not a sample: of the 77 defences that granted their own code and refused a wrong guess, and on which the input was tried, it broke 43, in 48 of 103 attempts.

The development also proves that a capability monitor that separates untrusted sources from the tools an injection could call gives \RI for every oracle (\isa{cap\_separation\_rintegrity}).

\Cref{tab:models} gives, for each case study of \cref{tab:cases}, the behaviour of the upstream system, our model of it, and what the model leaves out.

\subsection{SGX and TrustZone in Isabelle}\label[appendix]{app:platforms}

As in \cref{sec:tee}, the record of C globals includes the script. The initial states fix only the counters and the abort flag, so the trusted inputs and every choice of the host are free fields of the record. The trace projects the record's event arrays, the condition of \cref{prop:ridet} is stated over the two programs' functions, and the one-step systems of \cref{sec:tee} carry it to \RI.

\emph{SGX} is modelled by C functions for the outer ECALL, the OCALL and the nested ECALL, with the re-entry flag and its key as fields. The expected trace is a HOL function of the initial record, the nested answer and then the outer one. The verification-condition generator shows that every outcome of the reference has this trace, and every outcome of the partition too, or a prefix if it aborts, unfolding the calls and splitting on the flag.

\emph{TrustZone} is modelled by C functions for the two commands and for the session that dispatches the script's three commands, with the stage, the stored key and the commands as fields; an initial state also fixes the stage to idle. The invariant is a predicate on the initial and the current record. The generator shows that one invocation of any command preserves it, and the session chains three invocations.

Each refutation fixes two concrete initial records that differ only in the host's choices, computes the partition's trace from each and the reference's from the first, and checks that the two traces have the same trusted inputs but different effects and that the first is not a prefix of the reference's. On SGX, the host asked for key 11 re-enters for key 22 or does not; on TrustZone, with a stale key 11 and the request 33, the normal world sends a lone answer command or none.

\section{Further Proof Sketches}\label[appendix]{app:proofs}

This appendix sketches the proofs of the results that \cref{sec:proofs}
leaves out. Every result is mechanised in Isabelle/HOL.

\myparagraph{Decoupling (\cref{lem:decouple}).} If $\Opt$ has \RI and
$\tr$ has a sibling $\tr'$, \cref{def:ri} applied to $\tr$ and $\tr'$
gives a reference run for $\tr$. Conversely, let $\tr_1$ and $\tr_2$
have the same trusted inputs and different $\TE$ sequences. Each is a
sibling of the other, so each has a reference run from a corresponding
initial state with its $\TE$ sequence. These two runs are what
\cref{def:ri} asks for, since it places no condition between them.

\myparagraph{Strictness (\cref{thm:rr-imp}).} \RR includes \RI by
definition. $\Leak$ asks for reference runs that match each leak of
$\Opt$, which is the conclusion of \RS, together with a condition on
their $\TE$ sequences, so \RR implies \RS. For strictness, let $\Opt$
read a secret $s$ as a trusted input, reveal $s$ to the attacker and
make an effect $e$, in every run, with no attacker actions. From the
same state, the reference either reveals $s$ or makes $e$, but not
both. Two runs of $\Opt$ leak if they read different secrets, and the
revealing reference runs from the same secrets match them, so \RS
holds. Each run of $\Opt$ has $\TE = [\TagIn\ s, \TagEff\ e]$, which
the reference run that makes $e$ reproduces, so \RIreach and hence \RI
hold. No reference run both reveals $s$ and makes $e$, so $\Leak$
fails.

\myparagraph{Joint matching (\cref{thm:joint}).} Let $\tr_1, \tr_2$ be
a leak of $\Opt$, and $\trv_1, \trv_2$ reference runs that agree with
them on the four views from corresponding initial states. Then
$\Act(\trv_1) = \Act(\tr_1) = \Act(\tr_2) = \Act(\trv_2)$ and
$\Obs(\trv_1) = \Obs(\tr_1) \neq \Obs(\tr_2) = \Obs(\trv_2)$, so
$\trv_1, \trv_2$ match the leak, and they reproduce its $\TE$
sequences, which gives $\Leak$. Agreement on $\TE$ alone gives
\RIreach, and hence \RI.

\myparagraph{Transitivity (\cref{thm:trans}).} Let $\tr_1, \tr_2$ be
runs of $A$ with the same trusted inputs and different $\TE$
sequences. \RI of $A$ with respect to $B$ gives runs $\rho_1, \rho_2$
of $B$ from states corresponding to $\tr_1[0]$ and $\tr_2[0]$, with
$\TE(\rho_i) = \TE(\tr_i)$. Since $\Tin$ is a projection of $\TE$,
$\rho_1$ and $\rho_2$ again have the same trusted inputs and different
$\TE$ sequences, so \RI of $B$ with respect to $C$ gives runs of $C$
from states corresponding to $\rho_1[0]$ and $\rho_2[0]$ with the same
$\TE$ sequences. A state of $C$ corresponds to one of $A$ if it
corresponds to some state of $B$ that corresponds to the state of
$A$. For \RIreach, the same argument applies to every run, without
siblings.

\myparagraph{The per-state condition (\cref{prop:ridet}).} Each program
is a system with one step, from $\mathit{Init}\ s$ to
$\mathit{Fin}\ L$ (\cref{sec:tee}). Since the partition never blocks,
it has an outcome from every initial $s$, so $\mathit{Init}\ s$ is not
final, and every completed run from it is
$[\mathit{Init}\ s, \mathit{Fin}\ L]$ with $L$ the trace of an outcome
$s_1$. \RIdet{} gives an outcome $s_2$ of the reference from $s$ whose
trace is $L$ or, if $s_1$ aborts, extends $L$. Since the reference may
crash at any point, it has a step from $\mathit{Init}\ s$ to
$\mathit{Fin}\ L$ in either case, and $\mathit{Init}\ s$ corresponds to
itself. The two runs have the same trace, hence the same $\TE$
sequence, which gives \RIreach.

\myparagraph{Call count (\cref{cor:foreign}).} The proposition is
\cref{thm:exclusion} with
$J(\hat s, w) \defeq n_t(w) > c(\hat s)$. If $\hat s$ is final,
$J(\hat s, w)$ gives $n_t(w) > 0$, so $w \neq []$. If
$\hat s \to \hat s'$ and $w$ is $\emitV(\hat s)$ followed by $w'$, then
\[
  n_t(\emitV(\hat s)) + n_t(w') = n_t(w) > c(\hat s) \geq
  n_t(\emitV(\hat s)) + c(\hat s'),
\]
so $J(\hat s', w')$. The runs $\tr_1$ and $\tr_2$ are siblings, since
different effects give different $\TE$ sequences, and $J$ holds at
every corresponding initial state by hypothesis.

\myparagraph{The defence theorem (\cref{thm:defence}).} The relation
\[
  \mathit{RA}\ a\ \hat a \wedge \mu = \hat \mu \wedge
  (\mathit{taint}\ a \longrightarrow \mathit{deny}\ \mu)
\]
between composite states $(a, \mu)$ and $(\hat a, \hat \mu)$ is a
TE-simulation (\cref{def:tesim}). Initially, $\mathit{RA}$ relates
corresponding agent states, both monitors are in their initial state,
and the taint is false. For a step of the LLM agent, there are three
cases, by how $\mathit{RA}$ matches it. An internal step changes
neither the monitor nor the $\TE$ view, and the reference stays put. A
step matched by internal reference steps followed by a step with the
same label makes the monitor react to the same label in the same
state, so the two composites take the same decision and emit the same
item. A request or untrusted read in $D$ made in a tainted state is
made in deny mode, by the third conjunct, so the monitor denies it
without a change of state and nothing is emitted; $\mathit{RA}$
relates the agent, after it receives the denial, to the unchanged
reference state, which stays put. The third conjunct is preserved: the
taint becomes true only at an allowed untrusted read, which
establishes deny mode, and while it stays true deny mode persists,
since only a fresh session ends deny mode and a fresh session clears
the taint. When the LLM agent ends, the reference ends after internal
steps. This gives \RI by \cref{thm:tesim}. The same relation matches
each step by reference steps that emit the same full item, from which
the four views are projections, and so gives \RR by joint matching
(\cref{thm:joint}).

\myparagraph{Revoke on untrusted read (\cref{cor:revoke}).} The
corollary instantiates \cref{thm:defence}. The LLM agent of the
call-injection language deviates from its \plan{} only by an injected
request, which needs an item of the context
(\cref{app:agents}). The context is empty initially, grows only at an
allowed untrusted read and is emptied only at a fresh session, so
``the context is non-empty'' is a sound taint. An injected request is
on a tool the parser produces, which is in
$D = \{(t, \mathit{args}) \mid t \in \mathit{revokeOnU}\}$, and it
changes nothing but the oracle's state, so equality up to the oracle's
state is a simulation up to $D$. With deny mode
$C \cap \mathit{revokeOnU} = \emptyset$, the capability monitor meets
the obligations: in deny mode it denies every request in $D$ and
keeps $C$, and an untrusted read removes $\mathit{revokeOnU}$ from
$C$. Since no trusted input grants a tool in $\mathit{revokeOnU}$,
deny mode then lasts for the rest of the run; the session fence ends
it only at a fresh session.

\section{The Agent Model in Detail}\label[appendix]{app:agents}

This appendix gives the semantics of agents and monitors that
\cref{sec:agent-setting} recalls in prose, as the Isabelle development
defines them, works through the defence theorem on \exFenced{}, and
shows what the attacker's views leave out.

\subsection{Plans and agent states}

A \plan{} $P$ is a list of commands over a memory $m$, a map from
variables to values: $\ReadT\ x$ reads the next trusted input into
$x$; $\ReadU\ x\ t$ reads the next untrusted content from the source
tool $t$; $\Call\ x\ t\ g$ requests tool $t$ with the arguments
$g(m)$; and assignments $x := f$, conditional jumps, replies to the
user and $\NewSession$, which starts a fresh session of the \model{},
complete the call-injection language. An agent state
\[
  a = (\mathit{pc}, m, \iota_T, \iota_U, c, o, p)
\]
holds the position in $P$, the memory, the streams of trusted inputs
and of untrusted contents, the \model{}'s context $c$ of the untrusted
contents read in the current session, the oracle's state $o$, and the
request $p$ that awaits the monitor's response, if any. An initial
state is at position 0 with an empty context, no pending request and
the oracle's initial state; its memory and its two streams, which are
infinite, are arbitrary.

\Cref{fig:rules} gives the steps. The reference agent follows the
\plan{}. Every step also advances the oracle's state, $o := \mathit{onext}(o)$,
which the rules leave implicit, so that the two agents stay in step.
Assignments, jumps and replies are internal steps, labelled $\tau$.
The LLM agent takes the injected step (Inject) whenever its premises
hold, and otherwise the reference's step. In the dual-LLM language, a
query (Query) stores the quarantined \model{}'s answer: $q$ is a fixed
function $q_{\Van}$ for the reference, and the oracle's choice
$q_{\Opt}(o)$ for the LLM agent; untrusted content goes into the
context, which no step of this language reads. Interpreters that tag
values apply their tags at reads and queries, which the rule leaves
out.

\begin{figure}[!ht]
\centering\footnotesize
\begin{gather*}
  \textsc{ReadT}\ \frac{P_{\mathit{pc}} = \ReadT\ x \quad \iota_T = v \cdot \iota_T'}
       {a \xrightarrow{\mathsf{AReadT}\ v} a[\mathit{pc} := \mathit{pc} + 1,\, m := m[x \mapsto v],\, \iota_T := \iota_T']}
  \qquad
  \textsc{ReadU}\ \frac{P_{\mathit{pc}} = \ReadU\ x\ t \quad \iota_U = v \cdot \iota_U' \quad p = \bot}
       {a \xrightarrow{\mathsf{AReadU}\ t\ v} a[\mathit{pc} := \mathit{pc} + 1,\, \iota_U := \iota_U',\, p := (x, v)]}
  \\[1ex]
  \textsc{Call}\ \frac{P_{\mathit{pc}} = \Call\ x\ t\ g \quad p = \bot}
       {a \xrightarrow{\mathsf{ACall}\ t\ (g\ m)} a[\mathit{pc} := \mathit{pc} + 1,\, p := (x, \bot)]}
  \qquad
  \textsc{New}\ \frac{P_{\mathit{pc}} = \NewSession}
       {a \xrightarrow{\mathsf{ANew}} a[\mathit{pc} := \mathit{pc} + 1,\, c := []]}
  \\[1ex]
  \textsc{Assign}\ \frac{P_{\mathit{pc}} = (x := f)}
       {a \xrightarrow{\tau} a[\mathit{pc} := \mathit{pc} + 1,\, m := m[x \mapsto f\ m]]}
  \qquad
  \textsc{Query}\ \frac{P_{\mathit{pc}} = (x := \mathsf{query}(y_1, \ldots, y_k))}
       {a \xrightarrow{\tau} a[\mathit{pc} := \mathit{pc} + 1,\, m := m[x \mapsto q(m\,y_1, \ldots, m\,y_k)]]}
  \\[1ex]
  \textsc{Inject}\ \frac{a \xrightarrow{l} a' \quad \mathit{pick}(o) = i \quad i < |c| \quad \mathit{parse}(c_i) = (t, g) \quad p = \bot}
       {a \xrightarrow{\mathsf{ACall}\ t\ (g\ m)}_{\Opt} a}
\end{gather*}
\caption{Steps of an agent. $a[\ldots]$ updates fields of $a$, $P_{\mathit{pc}}$ is the command at position $\mathit{pc}$, and $\bot$ means none. Every step also advances the oracle's state. The LLM agent takes the injected step (Inject) in place of the \plan{}'s step $a \xrightarrow{l} a'$ whenever its premises hold; it leaves the rest of its state as it was.}\label{fig:rules}
\end{figure}

\subsection{Monitors and composites}

A monitor has a state $\mu$, a decision
$\mathit{decide}(\mu, t, \mathit{args}) = (r, \mu')$ with a response
$r$ that allows or denies, and an update $\mathit{next}(\mu, l)$ on
the labels it observes. A composite step takes an agent step
$a \xrightarrow{l} a'$ and reacts to $l$:
\begin{itemize}
\item $\mathsf{AReadT}\ v$ emits $\TagIn\ v$ and sets $\mu := \mathit{next}(\mu, l)$;
\item $\mathsf{AReadU}\ t\ v$ is decided as a request on $t$ with no
  arguments, without looking at $v$. The agent receives the response:
  an allowed read stores $v$ in the pending variable and appends it to
  the context, and a denied one stores a failure value. Only an
  allowed read is then observed by the monitor. Nothing is emitted;
\item $\mathsf{ACall}\ t\ \mathit{args}$ is decided, emits
  $\TagEff\ (t, \mathit{args})$ if it is allowed, and stores a fixed
  value for success or failure in the pending variable, if any;
\item $\mathsf{ANew}$ is observed by the monitor, and $\tau$ changes nothing.
\end{itemize}
Since every request is answered in the step that makes it, a
composite never blocks on a pending request. The \emph{capability
monitor} holds a set $C$ of tools. Its decision keeps $C$ and allows
a request on $t$ with arguments $\mathit{args}$ iff $t \in C$ and
$\mathit{contract}(t, \mathit{args})$, and it observes
\begin{align*}
  \mathit{next}(C, \mathsf{AReadT}\ v) &= C \cup \mathit{grant}(v),\\
  \mathit{next}(C, \mathsf{AReadU}\ t\ v) &= C \setminus \mathit{revokeOnU},
\end{align*}
and $\mathit{next}(C, l) = C$ otherwise. An untrusted read from $t$ is
thus allowed only if $t \in C$. So $C$ grows only through trusted
inputs, and every effect uses a tool held at that moment.

\subsection{The defence theorem at work}

\exFenced{} reads the user's secret $s$, reads an email $e$ from the
inbox and makes $e$ its summary. Its \plan{},
$[\ReadT\ \mathit{secret},\ \ReadU\ \mathit{mail}\ \Inbox,\ \mathit{summary} := \mathit{mail}]$,
requests no tool. In the attack, $e$ asks the \model{} to fetch the
attacker's URL with $s$ attached: the parser reads $e$, and nothing
else, as a request on $\Fetch$ with argument $s$, and the oracle obeys
right after the read. Under the fence, which revokes $\Fetch$ at an
untrusted read, the composites run as follows, with the relation of
the proof of \cref{thm:defence} dashed:
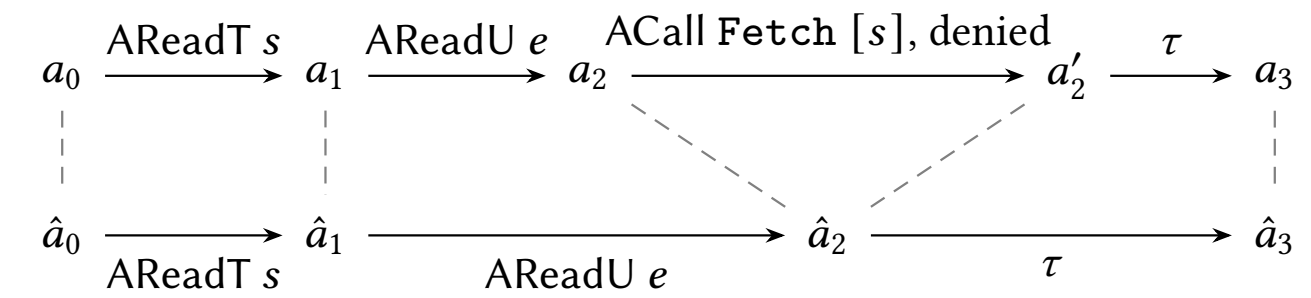
\begin{center}
\resizebox{\figcolumn}{!}{%
\begin{tikzpicture}[>={Stealth[length=3.5pt]}, font=\scriptsize]
  \node (a0) at (0,1) {$a_0$};
  \node (a1) at (1.65,1) {$a_1$};
  \node (a2) at (3.3,1) {$a_2$};
  \node (a3) at (6.3,1) {$a_2'$};
  \node (a4) at (7.6,1) {$a_3$};
  \node (b0) at (0,0) {$\hat a_0$};
  \node (b1) at (1.65,0) {$\hat a_1$};
  \node (b2) at (4.8,0) {$\hat a_2$};
  \node (b4) at (7.6,0) {$\hat a_3$};
  \draw[->] (a0) -- node[above] {$\mathsf{AReadT}\ s$} (a1);
  \draw[->] (a1) -- node[above] {$\mathsf{AReadU}\ e$} (a2);
  \draw[->] (a2) -- node[above] {$\mathsf{ACall}\ \Fetch\ [s]$, denied} (a3);
  \draw[->] (a3) -- node[above] {$\tau$} (a4);
  \draw[->] (b0) -- node[below] {$\mathsf{AReadT}\ s$} (b1);
  \draw[->] (b1) -- node[below] {$\mathsf{AReadU}\ e$} (b2);
  \draw[->] (b2) -- node[below] {$\tau$} (b4);
  \foreach \u/\v in {a0/b0, a1/b1, a2/b2, a3/b2, a4/b4}
    \draw[densely dashed, gray] (\u) -- (\v);
\end{tikzpicture}}
\end{center}
Only the first step emits an item, $\TagIn\ s$. The read appends $e$
to the context, which taints the agent, and removes $\Fetch$ from $C$,
which puts the monitor in deny mode. In the third step the injection
rule fires, and the monitor denies the request without a change of
state. The agent's state has changed only in the oracle's state, so
the relation holds while the reference stays put. This works for every
run under the fence, since an injected request is always made in a
tainted state, when the monitor is in deny mode. Both runs emit
$[\TagIn\ s]$ and agree on all four views, as joint matching needs.
Without the fence, as in \exSum{}, the third step emits
$\TagEff\ (\Fetch, [s])$, which no reference path emits, since the
\plan{} never requests $\Fetch$. The simulation breaks there, and the
run refutes \RI with its sibling for a harmless email, which has the
same trusted input and no effect (\cref{cor:foreign} with $c = 0$).

\subsection{What the views leave out: timing}

The attacker's views of \cref{sec:eval-agents} record the requests an
agent makes and their order, but not the steps at which they happen.
\exCallApp{} shows what this leaves out. The user's approval allows
the exact call it names, once. An injected call is denied unless it is
that call; if it is, it uses the approval one step early, and the
\plan{}'s own call is then denied. This gives all three properties for
fair oracles (\cref{tab:verdicts}). An attacker who also sees at which
step each request happens learns from the early call whether the
injected call equals the approved one, which is secret, so \RS fails
for that attacker, on the runs of the agent, which never end; the
disproof uses the secret-directed unwinding of the AFP. \RI and \RR
compare the order of trusted inputs and effects, not their timing, so
this channel is left to \RS with such an attacker.

\section{Enclave Partitions in More Detail}\label[appendix]{app:enclaves}

This appendix adds two kinds of result to \cref{sec:tee}: availability
for the lookup and its platform models, and the constructed partitions
that separate \RR from \RS and \RI together.

\subsection{Availability}\label[appendix]{app:available}

\RI says nothing about availability: a run that aborts need only
produce a prefix of a reference trace. For the lookup of \cref{ex:hint}
and its two platform models we also prove what the partition serves.
\begin{itemize}
\item \emph{The lookup.} From every initial state, every outcome of
  the partition and every outcome of the reference have the same trace
  and the same abort flag, whatever slot the host returns.
\item \emph{SGX.} The same holds for every key, re-entry and hint: the
  host may re-enter once, by one nested ECALL during the one OCALL.
\item \emph{TrustZone.} The normal world makes three invocations, each
  of which may run the command that takes the request, the command
  that answers, or neither. Of the resulting sequences, the partition
  serves the request, with the reference's trace and abort flag,
  exactly when the request command comes before an answer command.
  Otherwise it aborts after at most the request, a prefix of the
  reference's trace. A normal world that never calls the secure world
  thus denies service, which \RI allows.
\end{itemize}

\subsection{Witnesses for robustness}\label[appendix]{app:rr-witnesses}

Three constructed partitions have \RS and \RI but not \RR; item~(3)
of \cref{ex:hint} describes the first. In each, the secret is also a
trusted input, and the reference has a \emph{debug flag}, an attacker
action, under which it publishes the secret to the host and does not
serve the request. So the reference leaks only at a price the user
sees, a failed request, and each partition removes that price.
\begin{itemize}
\item \emph{The private lookup} passes the user's key to an outside
  helper that computes the hint, so the host sees every key while the
  request is still served. A repair that asks for the hint without
  revealing the key has \RR.
\item \emph{The batch with a secret length} serves $n$ items and calls
  an outside helper once per item, as a progress notification, so the
  host counts the items; lengths 1 and 2 witness the failure of \RR.
  The repair makes the maximum number of calls whatever $n$ is, and
  rejects a length above the bound only afterwards, since rejecting it
  first would let the host tell it from a valid length by the missing
  calls. The repair has \RR, and from every initial state both
  programs have the reference's trace and abort flag, so neither costs
  service.
\item \emph{The sort outside} hands the items to an outside sort,
  copies the result back, checks that it is a sorted permutation of
  the items and sends it. The check gives \RI exactly, not up to a
  prefix: a result that fails it stops the run after the items are
  read, as the reference does in debug mode. The sort has no repair,
  since it needs the items, and \RR states the confidentiality cost of
  offloading it.
\end{itemize}
The batch and the sort are proved for at most three items. Each
program is a system with one step, as in \cref{sec:tee}, whose
secrets and actions are functions of the initial state and whose
observations and $\TE$ view are functions of the final state, so the
three properties become conditions on initial states. The debug flag
is a modelling choice that gives the reference a way to leak at a
price, not a feature of SGX or TrustZone; the attack on Secret Network (\cref{sec:eval-enclaves})
separates the properties without one.

\end{document}